\documentclass[
  journal=pasa,
  manuscript=research-paper,
  year=2020,
  volume=37,
]{cup-journal}

\usepackage{amsmath}
\usepackage[nopatch]{microtype}
\usepackage{booktabs}
\usepackage{tabularx}
\usepackage{setspace,hyperref}
\hypersetup{colorlinks=true,
            linkcolor=blue,
            urlcolor=blue,
            linktoc=all,
            citecolor=blue}
\title{EMU/GAMA: A Technique for Detecting Active Galactic Nuclei in Low Mass Systems}
\author{Jahang Prathap}
\affiliation{School of Mathematical and Physical Sciences, Macquarie University, Sydney, NSW 2109, Australia}
\alsoaffiliation{Macquarie University Astrophysics and Space Technologies Research Centre, Sydney, NSW 2109, Australia}
\alsoaffiliation{ARC Centre of Excellence for All Sky Astrophysics in 3 Dimensions (ASTRO-3D), Australia}
\email[Jahang Prathap]{jahangprathap12@gmail.com}

\author{Andrew M. Hopkins}
\affiliation{School of Mathematical and Physical Sciences, Macquarie University, Sydney, NSW 2109, Australia}
\alsoaffiliation{Macquarie University Astrophysics and Space Technologies Research Centre, Sydney, NSW 2109, Australia}

\author{Aaron S. G. Robotham}
\affiliation{International Centre for Radio Astronomy Research (ICRAR), The University of Western Australia, M468, 35 Stirling Highway, Crawley, WA 6009, Australia}
\alsoaffiliation{ARC Centre of Excellence for All Sky Astrophysics in 3 Dimensions (ASTRO-3D), Australia}

\author{Sabine Bellstedt}
\affiliation{International Centre for Radio Astronomy Research (ICRAR), The University of Western Australia, M468, 35 Stirling Highway, Crawley, WA 6009, Australia}

\author{Jos\'e Afonso}
\affiliation{Instituto de Astrofísica e Ciências do Espaço, Universidade de Lisboa, OAL, Tapada da Ajuda, PT1349-018 Lisbon, Portugal}
\alsoaffiliation{Departamento de Física, Faculdade de Ciências, Universidade de Lisboa, Edifício C8, Campo Grande, PT1749-016 Lisbon}

\author{Ummee T. Ahmed}
\affiliation{Australian Astronomical Optics, Macquarie University, 105 Delhi Rd, North Ryde, NSW 2113, Australia}
\alsoaffiliation{Centre for Astrophysics, University of Southern Queensland, 37 Sinnathamby Boulevard, Springfield Central, QLD 4300, Australia}

\author{Maciej Bilicki}
\affiliation{Centre for Theoretical Physics, Polish Academy of Sciences, Al. Lotników 32/46, 02-668 Warsaw, Poland}

\author{Malcolm N. Bremer}
\affiliation{H.H. Wills Physics Laboratory, University of Bristol, Tyndall Avenue, BS8 1TL, Bristol, UK}

\author{Sarah Brough}
\affiliation{School of Physics, University of New South Wales, NSW 2052, Australia}

\author{Michael J. I. Brown}
\affiliation{School of Physics, Monash University, Clayton, VIC 3800, Australia}

\author{Yjan Gordon}
\affiliation{Physics Department, 2320 Chamberlin Hall, University of Wisconsin-Madison, 1150 University Avenue Madison, WI 53706-1390, USA}

\author{Benne W. Holwerda}
\affiliation{Department of Physics and Astronomy, University of Louisville, 40292 KY Louisville, USA}

\author{Denis Leahy}
\affiliation{Department of Physics and Astronomy, University of Calgary, Calgary, AB T2N 1N4, Canada}

\author{\'Angel R. \mbox{L\'opez-S\'anchez}}
\affiliation{School of Mathematical and Physical Sciences, Macquarie University, Sydney, NSW 2109, Australia}
\alsoaffiliation{Macquarie University Astrophysics and Space Technologies Research Centre, Sydney, NSW 2109, Australia}
\alsoaffiliation{ARC Centre of Excellence for All Sky Astrophysics in 3 Dimensions (ASTRO-3D), Australia}

\author{Joshua R. Marvil}
\affiliation{National Radio Astronomy Observatory, P.O. Box O, Socorro, NM 87801, USA}

\author{Tamal Mukherjee}
\affiliation{School of Mathematical and Physical Sciences, Macquarie University, Sydney, NSW 2109, Australia}
\alsoaffiliation{Macquarie University Astrophysics and Space Technologies Research Centre, Sydney, NSW 2109, Australia}
\alsoaffiliation{ARC Centre of Excellence for All Sky Astrophysics in 3 Dimensions (ASTRO-3D), Australia}

\author{Isabella Prandoni}
\affiliation{INAF – Istituto di Radioastronomia, Via P. Gobetti 101, 40129 Bologna, Italy}

\author{Stanislav S. Shabala}
\affiliation{School of Natural Sciences, University of Tasmania, Private Bag 37, Hobart, Tasmania 7001, Australia}

\author{Tessa Vernstrom}
\affiliation{International Centre for Radio Astronomy Research (ICRAR), The University of Western Australia, M468, 35 Stirling Highway, Crawley, WA 6009, Australia}

\author{Tayyaba Zafar}
\affiliation{School of Mathematical and Physical Sciences, Macquarie University, Sydney, NSW 2109, Australia}
\alsoaffiliation{Macquarie University Astrophysics and Space Technologies Research Centre, Sydney, NSW 2109, Australia}
\alsoaffiliation{ARC Centre of Excellence for All Sky Astrophysics in 3 Dimensions (ASTRO-3D), Australia}
%% First letter not capped
\begin{document}
\begin{abstract}
\noindent We propose a new method for identifying active galactic nuclei (AGN) in low mass ($\rm M_*\leq10^{10}M_\odot$) galaxies. This method relies on spectral energy distribution (SED) fitting to identify galaxies whose radio flux density has an excess over that expected from star formation alone. Combining data in the Galaxy and Mass Assembly (GAMA) G23 region from GAMA, Evolutionary Map of the Universe (EMU) early science observations, and Wide-field Infrared Survey Explorer (WISE), we compare this technique with a selection of different AGN diagnostics to explore the similarities and differences in AGN classification. We find that diagnostics based on optical and near-infrared criteria (the standard BPT diagram, the WISE colour criterion, and the mass-excitation, or MEx diagram) tend to favour detection of AGN in high mass, high luminosity systems, while the ``\textsc{ProSpect}'' SED fitting tool can identify AGN efficiently in low mass systems. We investigate an explanation for this result in the context of proportionally lower mass black holes in lower mass galaxies compared to higher mass galaxies and differing proportions of emission from AGN and star formation dominating the light at optical and infrared wavelengths as a function of galaxy stellar mass. We conclude that SED-derived AGN classification is an efficient approach to identify low mass hosts with low radio luminosity AGN. 
\end{abstract} 
\section{Introduction}
\noindent One of the primary objectives in extragalactic astronomy is the identification and differentiation of active galactic nuclei (AGN) from star forming galaxies (SFGs). A Significant number of galaxies host a supermassive black hole (SMBH), which acts as their central engine, and identification of these active nuclei is crucial in understanding their role in galaxy evolution \citep{1998AJ....115.2285M,2004ARA&A..42..603K,doi:10.1146/annurev-astro-082708-101811}.

\noindent Luminous radio sources ($\rm L_{1.4~GHz}>10^{24}~W/Hz$) are typically AGN, whereas only a few percent, typically around 10\%, are considered radio-loud \citep{2000ApJS..126..133W}. This fraction increases with optical and \mbox{X-ray} luminosities \citep{1995PASP..107..803U}. In the mid-infrared (mid-IR) regime, strong ionisation lines and polycyclic aromatic hydrocarbon (PAH) features serve as excellent diagnostics for identifying AGN \citep{2000A&A...359..887L}. Luminous, centrally-located compact \mbox{X-ray} and $\gamma$-ray sources are classified as AGN and do not require confirmation from other methods \citep[][]{2004ASSL..308...53M}. However, depending on its accretion state or mode and duty cycle, the presence of an active nucleus might not always be simultaneously visible in all spectral regions, and consequently, not all diagnostic methods are necessarily consistent \citep{1979MNRAS.188..111H,2007MNRAS.376.1849H}.

\noindent A significant fraction of radio sources do not show broad optical emission lines or even any emission lines at all (BL Lacertae objects; \citealt{1995MNRAS.277.1477P}, low-excitation radio galaxies (LERGs); \citealt{2012MNRAS.421.1569B}), adding complexity to the classification process. Infrared observations of both \mbox{X-ray} selected and optically selected AGN samples reveal a wide range of IR to X-ray SEDs and significant variation in the IR colours of AGN \citep[][]{2004ASSL..308...53M,2017A&ARv..25....2P}. Consequently, relying solely on IR colours can pose challenges in conducting an AGN survey. The IR spectral parameters do not exhibit a correlation with the optical spectral slope, IR luminosity, mid-IR-to-far-IR luminosity ratio, or the effects of inclination-dependent extinction in the context of a dusty torus model \citep{2004ASSL..308...53M}. Additionally, certain AGN classes display relatively low soft X-ray-to-optical ratios, indicating the existence of an IR-selected AGN group with broad optical emission lines that are considered X-ray ``weak'' \citep{2002ApJ...564L..65W}. The key takeaway is that a multiwavelength analysis is essential for accurately identifying galaxies hosting AGN.
\subsection{Classifying AGN}
\noindent Currently, a wide array of diagnostics is employed for AGN identification. A traditional optical diagnostic, the Baldwin-Phillips-Terlevich \citep[BPT;][]{1981PASP...93....5B} diagram, uses the intensity ratios of [OIII]/H$\beta$ and [NII]/H$\alpha$ to classify galaxies as AGN or SFGs. It was later revised by \cite{veilleux1987spectral} to include [SII] and [OI] lines, \cite{kewley2001optical} to provide a theoretical maximum for SF, and \cite{2003MNRAS.346.1055K} for a semi-empirical demarcation between SFGs and AGN making the optical selection more comprehensive. For higher redshifts ($z > 0.4$), where [NII] and H$\alpha$ lines may not be available, alternative diagnostic diagrams have been developed. The mass-excitation \citep[MEx;][]{2011ApJ...736..104J} diagnostic diagram uses the [OIII]/H$\beta$ line ratio and stellar mass as a substitute for [NII]/H$\alpha$, while the colour-excitation \citep[CEx;][]{2011ApJ...728...38Y} diagnostic diagram employs the rest-frame $U-B$ colour along with the [OIII]/H$\beta$ line ratio for classification. In addition to the BPT diagram, other diagrams use different line ratios to classify galaxies. The blue diagram \citep{2004MNRAS.350..396L} incorporates the [OIII]/H$\beta$ and [OII]/H$\beta$ line ratios, while the equivalent width (EW) of H$\alpha$ (W$_{H\alpha}$) versus [NII]/H$\alpha$ \citep[WHAN,][]{2010MNRAS.403.1036C,2011MNRAS.413.1687C} diagram uses the [NII]/H$\alpha$ line ratio combined with EW of H$\alpha$. The Trouille-Barger-Tremonti \citep[TBT,][]{2011ApJ...742...46T} diagram uses the rest-frame ($g-z$) colour and the line-ratio [NeIII]/[OII]. The OHNO diagram \citep{2015AAS...22520604Z,2022ApJ...926..161B} uses the line ratios [OIII]/H$\beta$ and [NeIII]/[OII] for classification.

\noindent The mid-IR region can be used to produce AGN diagnostics that primarily uses colours and line-flux intensities originating from dust re-emission. Several methods (the Lacy wedge; \citealt{2004ApJS..154..166L, 2007AJ....133..186L, 2005ApJ...621..256S}, the Stern wedge; \citealt{2005ApJ...631..163S}, the Infrared Array Camera (IRAC) power-law; \citealt{2006ApJ...640..167A, 2007ApJ...660..167D, 2012ApJ...748..142D}, and the KI \& KIM methods; \citealt{2012ApJ...754..120M, 2014A&A...562A.144M}) have been developed, each with its specific approach and advantages. To mitigate degeneracies, incorporating redshift information along with colour parameters has been suggested \citep{2013ApJ...764..176J}. Wide-field Infrared Survey Explorer \citep[WISE,][]{2010AJ....140.1868W, 2011ApJ...735..112J} photometry has also been used to derive criteria for distinguishing AGN from SFGs \citep{2012MNRAS.426.3271M,stern2012mid,2013ApJ...772...26A,2018ApJS..234...23A}. Mid-IR line ratios have been explored for AGN identification as well \citep{Genzel_1998,2002A&A...393..821S}. No AGN selection is 100\% complete and IR AGN selection is no exception. Several studies \citep[e.g.,][]{2006ApJ...642..126B,2014MNRAS.438.1149G,2012ApJ...754..120M} have highlighted the instances of contamination caused by high-z galaxies or inactive galaxies.

\subsection{Radio AGN}
\noindent One of the key properties of AGN is how their emission is distributed across the entire electromagnetic spectrum. The signature is not uniform throughout the spectrum as noted above, resulting in numerous ``types'' or classes of AGN depending on how they are diagnosed. In terms of radio properties, the AGN population can be broadly classified into radio-loud and radio-quiet depending on the radio power (e.g., radio-to-optical flux ratio; \citealt{1989AJ.....98.1195K}, radio luminosity; \citealt{1999MNRAS.309.1017W}). Since the accretion rate of AGN can still vary within these classes, based on the spectral intensities, AGN are further classified into radiative-mode and jet-mode \citep{1979MNRAS.188..111H,2007MNRAS.376.1849H,2014ARA&A..52..589H}. Radiative-mode AGN, where the energy is transferred radiatively from the accretion disk, consist of high-excitation radio galaxies \citep[HERGs,][]{2012MNRAS.421.1569B,2016MNRAS.460....2P} or strong line radio galaxies \citep[SLRGs,][]{2016A&ARv..24...10T}, Seyferts, and radio-quiet quasi-stellar objects (QSOs). Jet-mode AGN, whose primary energy output is in two-sided collimated jets, consist of low-excitation radio galaxies \citep[LERGs,][]{2012MNRAS.421.1569B,2016MNRAS.460....2P}, or weak line radio galaxies \citep[WLRGs,][]{2016A&ARv..24...10T} and LINERs \citep[when they are AGN,][]{2016MNRAS.461.3111B}.

\subsection{Spectral Energy Distributions and AGN}
\noindent The SED of a galaxy serves as a comprehensive blueprint, reflecting the intrinsic characteristics of a galaxy, including its stellar and gas content, age and enrichment, dust content, and AGN activity. SED fitting is, hence, a powerful technique that allows for derivation of the physical properties of a galaxy by fitting a model to the observed data \citep[e.g.,][]{2011Ap&SS.331....1W,2013ARA&A..51..393C}. SED fitting has also been used as an AGN diagnostic tool \citep{2022MNRAS.509.4940T,2023MNRAS.523.1729B,2023ApJ...950L...5Y}.

\noindent There are many state-of-the-art SED fitting tools that are commonly used (see \citealt{Pacifici_2023} for a short description of some of these tools and their application to datasets). We choose \textsc{ProSpect} \citep{2020MNRAS.495..905R,2020MNRAS.498.5581B,2022MNRAS.509.4940T} as a generative galaxy SED package to perform SED fitting. Our choice is motivated by the capability of \textsc{ProSpect} to extend the SED to the radio regime. Also, the dust modelling used in \textsc{ProSpect} uses the two-component form (see \S\,\ref{sec:diag_prospect}). Few SED fitting tools combine these two features. Here we explore the AGN diagnostic capabilities of the SED fitting technique through radio selection and find that this technique selects an AGN population residing in host galaxies of low stellar mass, which also show low radio luminosities.
  
\noindent The structure of the paper is as follows: We describe the photometric and spectroscopic data in \S\,\ref{sec:data}. The diagnostic tools are discussed in \S\,\ref{sec:diagnostics}, where we also introduce a binary classification scheme to identify which criteria classify each object as star forming or AGN. We compare different AGN diagnostic methods in \S\,\ref{sec:results}. We demonstrate the capability of \textsc{ProSpect} as a new tool for identifying AGN using SED fitting in low mass galaxies in \S\,\ref{sec:discussion}. We summarise our main findings and conclude in \S\,\ref{sec:conclusion}.
\noindent Throughout the paper we assume the \cite{2016A&A...594A..13P} cosmology with $\rm H_0 = 67.8~km~s^{-1}~Mpc^{-1}$, $\Omega_M = 0.308$ and $\Omega_\Lambda = 0.692$. The \textsc{ProSpect} tool adopts the \cite{Chabrier_2003} initial mass function (IMF).
\section{Data}\label{sec:data}
\noindent SED fitting requires panchromatic data. We fit SEDs to observed photometry spanning the far-ultraviolet (FUV) to the far-infrared (FIR) and predict the radio output. We then compare it with the observed radio emission. We use data from two main surveys: the Galaxy and Mass Assembly \citep[GAMA,][]{2009IAUS..254..469D,2011MNRAS.413..971D,2022MNRAS.513..439D,2015MNRAS.452.2087L} survey (FUV-FIR bands) and early science data from the Evolutionary Map of the Universe \citep[EMU,][]{2011PASA...28..215N, 2021PASA...38...46N,10.1093/mnras/stac880} survey (radio, at 888 MHz). Additionally, we incorporate data from the ALLWISE catalogue \citep{2012yCat.2311....0C,2014yCat.2328....0C}, which is derived from WISE for IR photometry.  

\subsection{GAMA Data}\label{sec:data_gama}
\noindent The GAMA Survey has emerged as a natural extension of legacy surveys such as the Sloan Digital Sky Survey \citep[SDSS,][]{2000AJ....120.1579Y} and the Two Micron All-Sky Survey \citep[2MASS,][]{2006AJ....131.1163S}. This comprehensive survey covers five separate fields, with a limiting magnitude of $\rm r < 19.8$ mag ($\rm i < 19.2$ mag for the G23 field), collectively spanning 286 square degrees. Four of the GAMA fields (G09, G12, G15, and G23) are strategically located within the footprints of the European Southern Observatory (ESO) VST Kilo Degree Survey \citep[KiDS,][]{2015A&A...582A..62D} and ESO Visible and Infrared Survey Telescope for Astronomy (VISTA) Kilo-degree Infrared Galaxy Public Survey \citep[VIKING,][]{2013Msngr.154...32E} surveys. These, along with data from Galaxy Evolution Explorer \cite[GALEX,][]{2005ApJ...619L...1M}, WISE, and Herschel \citep{2010A&A...518L...1P}, ensure that the GAMA survey benefits from extensive multiwavelength coverage from FUV to FIR. 

\noindent GAMA Data Release 4 \citep{2022MNRAS.513..439D} combines photometric data from these surveys together with the spectroscopy from the Anglo-Australian Telescope (AAT) AAOmega spectrograph. We are interested in the G23 field (338.1$^o$<RA<351.9$^o$, -35$^o$<$\delta$<-30$^o$) as it has radio measurements available from the EMU early science observations \citep{10.1093/mnras/stac880}.

\noindent The GAMA database is stored as tables in data management units (DMUs). The DR4 contains the \verb|gkvScienceCatv02| table \citep{2020MNRAS.496.3235B} in the \verb|gkvInputCatv02| DMU, with photometry for 458,844 sources and the \verb|gkvFarIRv03| table \citep{2020MNRAS.496.3235B} in the \verb|gkvFarIRv03| DMU with photometry for 458,783 sources. Spectra are taken from the \verb|GaussFitSimplev05| (\citealt[][49,623 sources]{2017MNRAS.465.2671G}) in the \verb|SpecLineSFRv05| DMU and stellar mass estimates from \verb|StellarMassesGKVv24| (\citealt[][67,006 sources]{2011MNRAS.418.1587T}) in the \verb|StellarMassesv24| DMU. We use Topcat \citep{2005ASPC..347...29T} to cross-match between these tables, forming the `GAMA' sample of 47,774 sources with the spectrophotometric data.
\subsection{EMU Data}\label{sec:data_radio}
\noindent The Australian Square Kilometre Array Pathfinder (ASKAP, \citealt{2007PASA...24..174J, mcconnell2016australian, hotan2021australian}) is a precursor telescope for the Square Kilometre Array (SKA). ASKAP is designed to observe in the frequency range of 800–1800 MHz with an instantaneous bandwidth of 288 MHz and offers rapid observations with good resolution and sensitivity.

\noindent EMU \citep{2011PASA...28..215N,2021PASA...38...46N} is an ongoing wide-field radio continuum survey conducted using ASKAP. Its primary goal is to provide a deep radio continuum map of the entire southern sky, extending up to +30$^o$ in declination. The survey achieves a sensitivity of approximately 20 $\mu$Jy/beam and a resolution of 15". As the survey progresses, it is anticipated that EMU will detect and catalogue approximately 40 million galaxies, including SFGs up to redshift $z\sim1$, powerful starbursts reaching even greater redshifts, and AGN to the edge of the visible Universe.

\noindent The G23 field has been observed as part of the ASKAP commissioning \citep{leahy2019askap} and later as part of the EMU early science program. In this study, we use the most recent ASKAP observation of the G23 field, carried out as part of EMU early science \citep{10.1093/mnras/stac880}. These data cover an area of 82.7 deg$^2$ and achieve a sensitivity of 0.038 mJy/beam at a frequency of 887.5 MHz. 55,247 sources were detected in this field. Of these, 39,812 have signal-to-noise (S/N) ratios $\geq$ 5. 

\noindent In our analysis, we focus on the common sources between GAMA and EMU early science radio data in G23. A cross-match between the two employing a matching radius of 5" \citep{2021PASA...38...46N,2023arXiv231211883A} results in 10,991 common sources with FUV-FIR photometry and 888 MHz radio measurements. Of these, there are 6425 sources with spectroscopic measurements and stellar masses. We refer to this sample as `GAMA-EMU.' We adopt the integrated radio flux densities from \citet{10.1093/mnras/stac880} for our analysis throughout, to ensure we are using the total radio flux densities for each source.
\subsection{WISE Data}\label{sec:data_wise}
\noindent WISE \citep[][]{2010AJ....140.1868W, 2011ApJ...735..112J} is a NASA medium-class explorer mission consisting of a 40 cm space-based infrared telescope that conducted a comprehensive digital imaging survey of the entire sky. WISE observations were performed in four mid-infrared bandpasses, namely 3.4, 4.6, 12, and 22 micrometres, denoted as $\rm W_1, W_2, W_3,~and~W_4$ with minimum 5$\sigma$ sensitivities of 0.08, 0.11, 0.8, and 6 mJy, respectively. During its mission, WISE produced and released a digital image atlas, providing coverage of the sky in four survey bands.

\noindent The ALLWISE catalogue \citep{2012yCat.2311....0C,2014yCat.2328....0C} is an extensive database containing $\rm W_1$ through $\rm W_4$ magnitudes of 747,634,026 sources. Cross-matching the ALLWISE catalogue with GAMA produced 44,244 sources within a matching radius of 3", which we refer to as `GAMA-WISE' throughout the paper. The cross-match radius is motivated by the fact that the point spread function (PSF) fitting of the ALLWISE pipeline can affect the resulting accuracy of the position particularly for resolved sources \citep[][]{2014ApJ...782...90C}. The latest data release of WISE \cite[the CatWISE catalogue,][]{2021ApJS..253....8M} provides $\rm W_1$ and $\rm W_2$ magnitudes of more than a billion objects with improved photometry. The choice of ALLWISE over CatWISE in this analysis is based on the diagnostic in \S\,\ref{sec:diag_wise}, which is defined based on ALLWISE photometry. Tests using the CatWISE measurements show no significant differences in our main results, and we adopt the ALLWISE measurement for consistency throughout. In order to link this to the galaxies with radio measurements, we cross-match the GAMA-WISE sample with the GAMA-EMU sample. This resulted in 6266 sources with FUV-FIR photometry, WISE magnitudes, and radio measurements, which we refer to as `GAMA-WISE-EMU.'
\subsection{Final Sample}\label{sec:data_final}
\noindent We now have 6266 galaxies with FUV-radio photometry, mid-IR magnitudes, relevant spectroscopic data and stellar mass estimates. We selectively remove galaxies under the following categories:
\begin{itemize}
    \item those with redshifts below the quality index (denoted as nQ in the \verb|GaussFitSimplev05| catalogue), $\rm nQ<3$ \citep{2017MNRAS.465.2671G}; 123 sources
    \item those with spectral lines in absorption, including H$\alpha$ and H$\beta$ EWs; 2330 sources
    \item those with unphysical stellar masses; 2 sources
    \item those with a starmask \citep[potential contamination of the photometry of a galaxy by a background star,][]{2020MNRAS.496.3235B}: 149 sources.
\end{itemize}
\noindent We are now left with 3662 sources. We do not exclude galaxies further on the basis of the Balmer line S/N ratio so as not to further reduce the sample. If we impose a $\rm S/N > 2$ limit, we are left with only 2253 galaxies, and our main results remain unchanged. The requirement to have H$\alpha$ lines for the BPT classification limits our sample to $z<0.34$.
\begin{table}[hbt!]
    \centering
    \begin{tabular}{c|c}
    \hline
        Catalogue & Number of Sources \\
        \hline
        GAMA & 47774 \\
        GAMA-WISE & 44244 \\
        GAMA-EMU & 6425 \\
        GAMA-WISE-EMU & 6266 \\
        GEM & 2956 \\ 
    \end{tabular}
    \caption{The number of galaxies cross-matched between GAMA G23, WISE, and EMU.}
    \label{tab:sample}
\end{table}

\noindent In our analysis of the \textsc{ProSpect} SED outputs described in detail below in \S\,\ref{sec:diag_prospect}, we remove the bad fits by using the log-posterior (LP) parameter, which serves as a goodness-of-fit metric akin to the $\chi^2$ statistic. The LP values within our sample exhibit a wide range, spanning -15 to -4000, with the more negative values standing out as outliers. Following a careful examination, we set a threshold of LP $\geq$-50. We are left with a sample of 2958 galaxies. To further validate the quality of this sample, each of these 2958 galaxies was visually inspected, resulting in the removal of two bad fits. This gives our final sample of 2956 galaxies, which we call `GEM' (for the final robustly defined GAMA-WISE-EMU sample) throughout the paper. A summary of the catalogues is shown in table \ref{tab:sample}.
\section{AGN Diagnostics}\label{sec:diagnostics}
\noindent We introduce a new AGN diagnostic based on the \textsc{ProSpect} SED fitting tool. In order to understand how this new diagnostic relates to existing well established methods, we use a selection of these here. We choose to compare against the BPT diagnostic \citep{1981PASP...93....5B}, the WISE $\rm W_1-W_2$ criterion \citep{2018ApJS..234...23A}, and the mass-excitation (MEx) diagram \citep{2011ApJ...736..104J}. There is an extensive array of AGN diagnostics in the literature, but of necessity, we limit our choice here to a selection of the more common in order to keep the analysis tractable. The GEM sample is selected in such a way that all galaxies have the relevant parameters to be classified by this suite of diagnostics.
\begin{figure}[h!]
    \centering
    \includegraphics[width=0.98\linewidth]{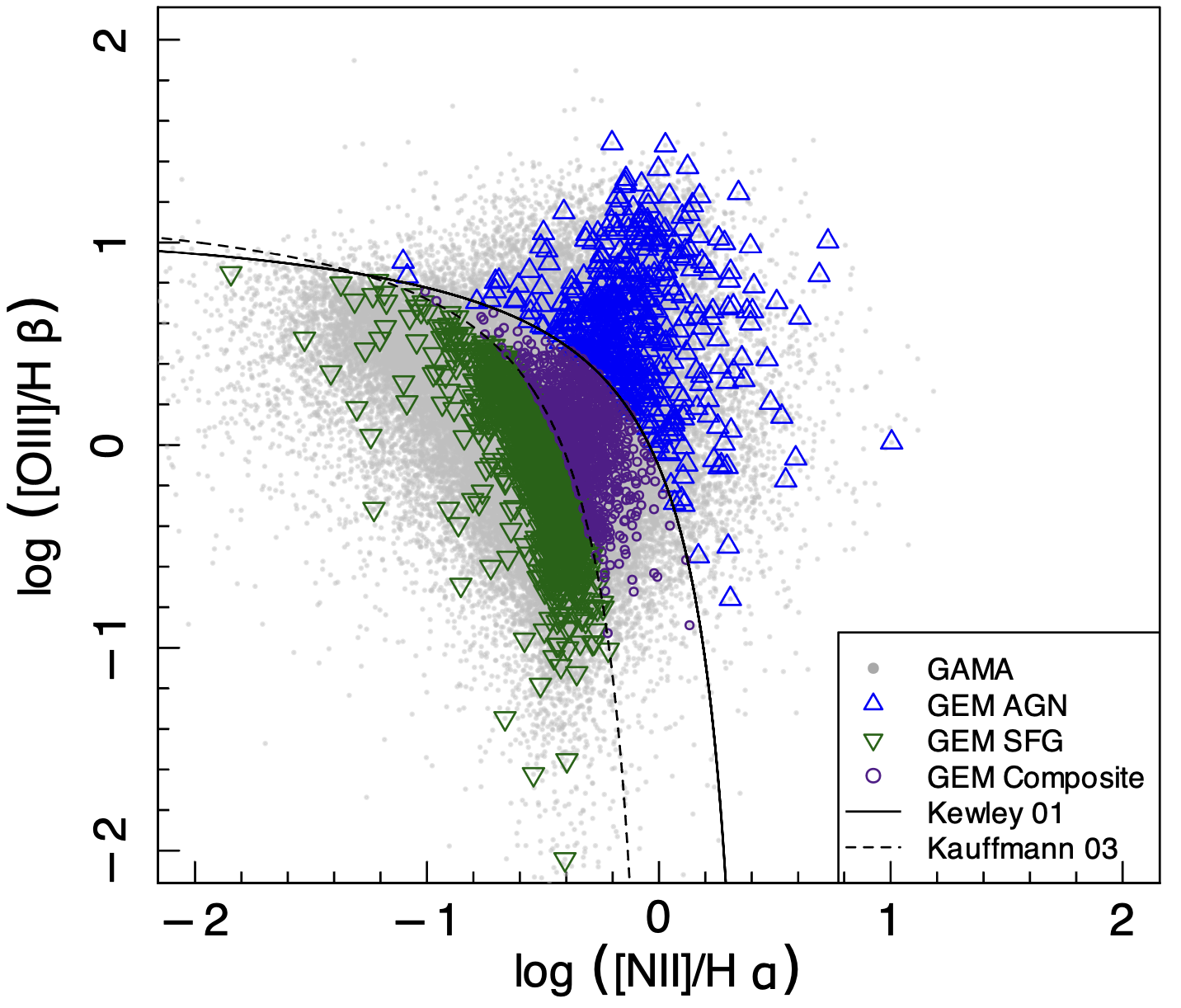}
    \caption{The BPT diagram: an emission line diagnostic diagram which uses the ratios [OIII]/H$\beta$ and [NII]/H$\alpha$ to delineate SFGs from AGN. The diagram shows the data from the full survey (G09, G12, G15, and G23) as GAMA BPT SFGs and AGN (grey dots),  GEM BPT SFGs (green triangles), and GEM BPT AGN (blue triangles). Galaxies below the theoretical Kewley line (solid line) and above the empirical Kauffmann line (dashed line) form a sequence of composite galaxies (purple dots). These galaxies are included among the BPT AGN in our analysis.}
    \label{GAMA_BPT}
\end{figure}

\noindent Balmer lines are affected by stellar absorption. To account for this, we apply a simple constant EW correction of 1.3 \AA~(\citealt{2003ApJ...599..971H,2011MNRAS.415.1647G}). It is important to note that the spectra in the GAMA dataset that are drawn from the Two-degree Field Galaxy Redshift Survey (2dFGRS) survey are not flux calibrated. For diagnostic diagrams, however, since we use ratios of lines close in wavelength, the lack of flux calibration is not an issue. Whenever Balmer line luminosities are used, we address both the stellar absorption and the flux calibration issues. To achieve this, we use the measured EWs and the r-band absolute magnitude to derive the respective luminosities, following the method outlined in \cite{2003ApJ...599..971H}.
\subsection{The BPT Diagram}\label{sec:diag_bpt}
\noindent In the seminal work by \cite{searle1971evidence}, it was demonstrated that emission-line spectra from different types of HII regions can be classified based on line-intensity ratios, such as [OIII]$\lambda$5007/H$\beta$ (excitation) or [OII]$\lambda$3727/[OIII]$\lambda$5007. However, relying solely on the excitation threshold, though powerful for identifying Seyfert 2 galaxies, has some drawbacks. It is prone to contamination by sufficiently ionised starbursts and often excludes many LINERs \citep{1981PASP...93....5B,osterbrock1985optical}. It was also found that the strength of low-ionisation lines such as [NII], [SII], and [OI] can be used to distinguish HII regions from Seyfert 2s \citep{1981PASP...93....5B,osterbrock1985optical}.

\noindent To address these limitations, \cite{1981PASP...93....5B} extended the one-parameter model to a more comprehensive two dimensional classifier of HII regions resulting from power-law photoionisation and shock wave heating. Further studies on a more accurate classification of HII regions, including a broader range of sources like Seyferts, LINERs, and narrow-line radio galaxies (NLRGs) followed. 

\noindent Notably, \cite{osterbrock1985optical} and \cite{veilleux1987spectral} provided a semi-empirical boundary line that effectively separates starbursts from Seyfert galaxies.  
\begin{figure}[h!]
    \centering
    \includegraphics[width=0.92\linewidth]{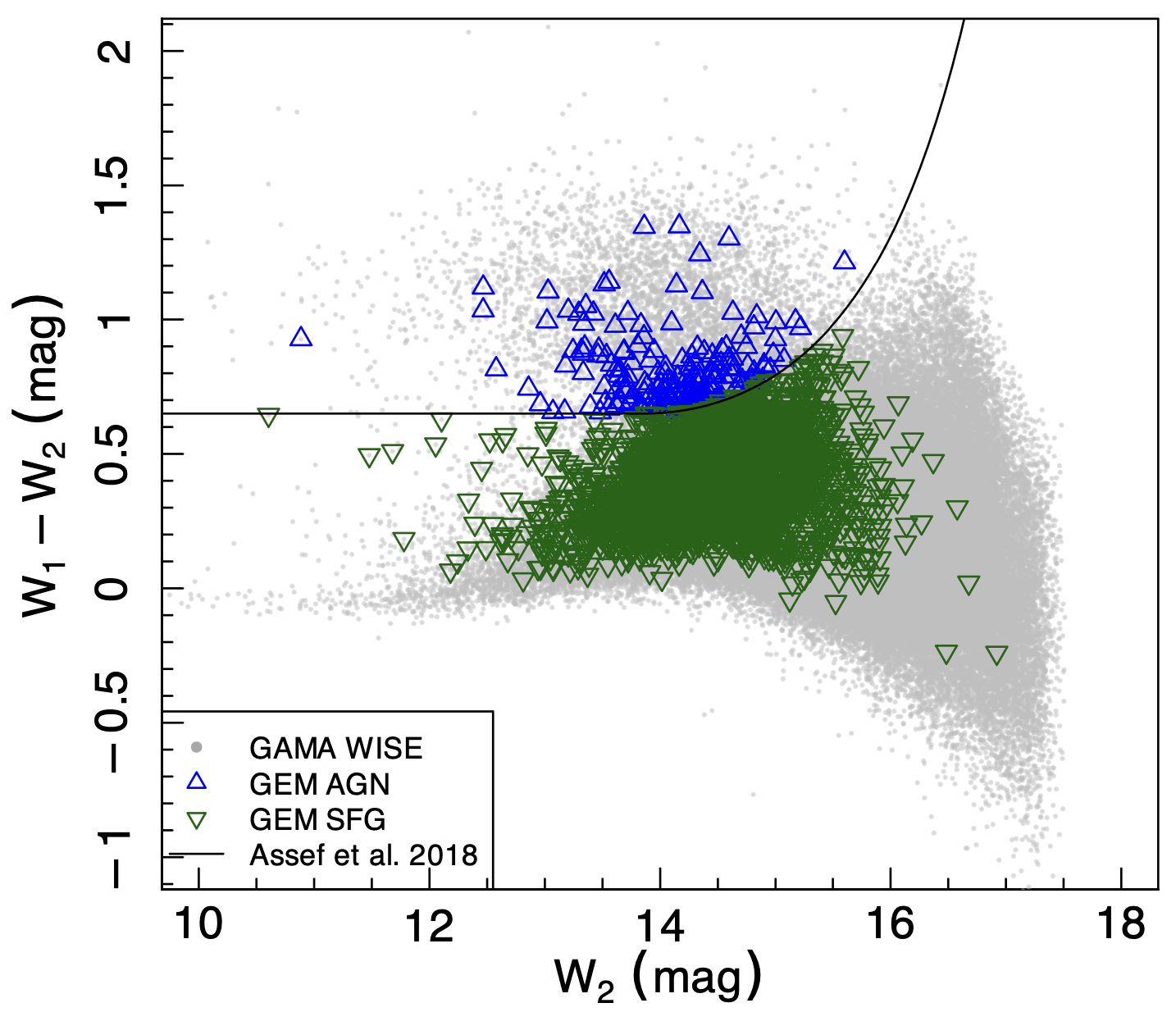}
    \caption{WISE colour diagram. The WISE AGN selection criterion described in \S\,\ref{eq:wise}, is shown as the solid line. The diagram shows the data from the full survey (G09, G12, G15, and G23) as GAMA WISE SFGs and AGN (grey dots), GEM WISE SFGs (green  triangles), and GEM WISE AGN (blue  triangles).}
    \label{GAMA_WISE}
\end{figure}
\noindent In our study, we employ the BPT diagram with the line ratios [OIII]/H$\beta$ versus [NII]/H$\alpha$, and we use the theoretical boundary line proposed by \cite{kewley2001optical} and the empirical line proposed by \cite{2003MNRAS.346.1055K} to distinguish between AGN and SFGs. Figure \ref{GAMA_BPT} illustrates the BPT classifier showing the GAMA data and our GEM sample. The data points in blue and green represent the GEM AGN and SFGs, respectively, overlaid on the GAMA sample shown in grey. Following the prescription in \cite{2003MNRAS.346.1055K}, we select all those galaxies above the dashed line as AGN. Composite galaxies, shown here in purple dots, are incorporated into our AGN classification for all our subsequent analyses unless otherwise noted in order that we are as complete as possible in identifying contributions from an AGN component.
\subsection{The WISE criterion}\label{sec:diag_wise}
\noindent AGN derive their power from the accretion disk formed by matter falling into the SMBH. The dusty torus surrounding most of the AGN absorbs a part of this power and re-emits this in the IR. This signature is evident across the IR and contributes measurably to the mid-IR spectrum. This process gives AGN characteristic red mid-IR colours. As a result, mid-IR colour criteria are a good way to select AGN \citep{stern2012mid}. Consequently, WISE data are now extensively used for AGN selection.
\begin{figure}[h!]
    \centering
    \includegraphics[width=0.95\linewidth]{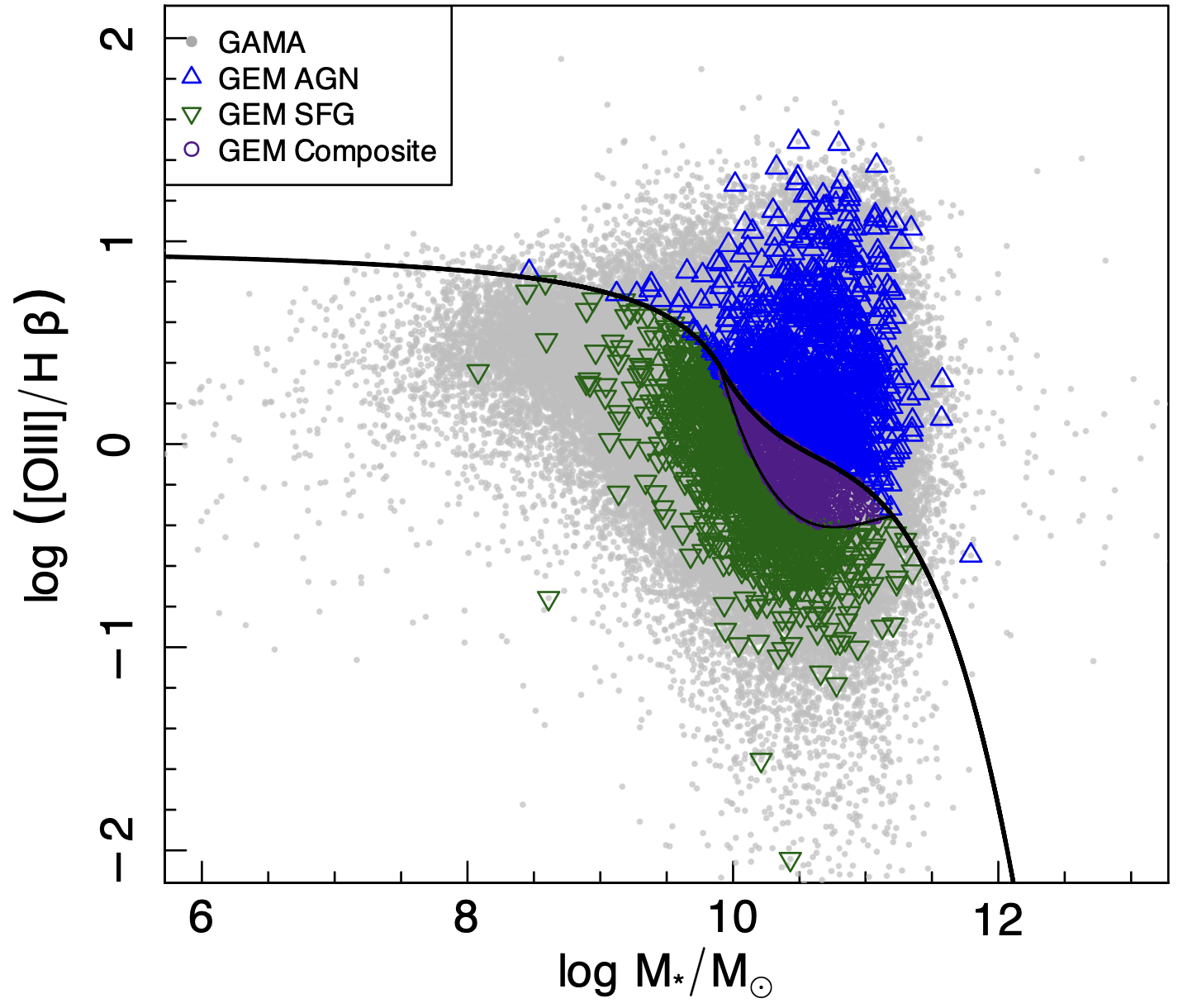}
    \caption{The mass-excitation (MEx) diagnostic. This diagnostic compares the excitation ([OIII]/H$\beta$) and stellar mass. The diagram shows the data from the full survey (G09, G12, G15, and G23) as GAMA MEx SFGs and AGN (grey dots), GEM MEx SFG (green  triangles), GEM MEx composite (purple dots), and GEM MEx AGN (blue  triangles) are shown. The regions occupied by SFGs, composite and AGN in the MEx parameter space are analogous to those in the BPT parameter space. The MEx composites are included among MEx AGN in our analysis.}
    \label{fig:MEx}
\end{figure}
\begin{figure*}[h!]
    \centering
    \includegraphics[width=0.95\linewidth]{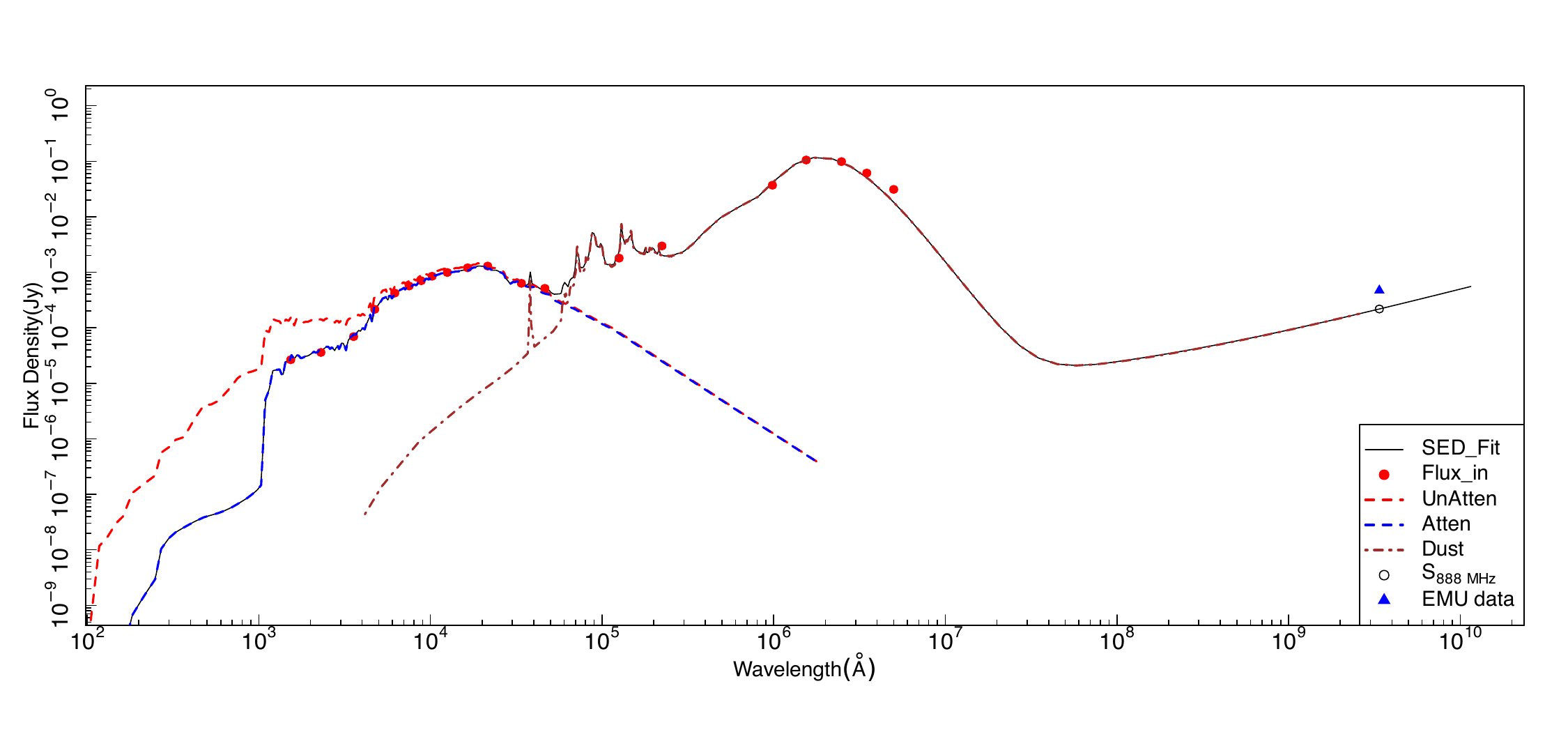}
    \caption{An example  SED fit using \textsc{ProSpect} (black solid line). The red data are the photometry from the GAMA catalogue. The EMU observation is shown as the blue triangle and the \textsc{ProSpect} predicted radio flux is shown as the black open circle. The unattenuated stellar emission (red dashed line) is subjected to dust absorption and the resulting attenuated stellar emission is shown as the blue dashed line. The corresponding energy difference is re-emitted in the infrared spectrum (brown dot-dashed line) taking energy balance into account. The ProSpect AGN component is not used here, thus the radio prediction arises from star formation processes alone. Based on our criterion, this example is classified as a \textsc{ProSpect} SFG.}
    \label{fig:SED}
\end{figure*}

\noindent \cite{stern2012mid} put forward a simple magnitude independent colour criterion, $\rm W_1-W_2 \geq 0.8$, providing a reliable AGN sample for $\rm W_2<15.05$ mag. As an extension to fainter magnitudes, \cite{2013ApJ...772...26A} developed a new mid-IR magnitude-dependent selection criterion. There are four flavours of the criterion depending on the level of reliability and completeness of AGN selection for a limiting magnitude, $\rm W_2 = 17.11$ mag. Here, we adopt the magnitude-dependent AGN selection criterion described in \cite{2018ApJS..234...23A} developed for ALLWISE data:
\[
   W_1 - W_2> 
\begin{cases}
    \alpha_R~\exp^{\beta_R(W_2 - \gamma_R)^2}&, W_2>\gamma_R\\
    \alpha_R              &, W_2 \leq \gamma_R,
    \label{eq:wise}
\end{cases}
\]
optimised for reliability, where the coefficients, ($\alpha_R, \beta_R, \gamma_R$) = (0.650, 0.153, 13.86), correspond to 90\% reliable AGN selection. Any other choice of WISE colour diagnostic, for instance, the $\rm W_1-W_2$ classifier \citep{stern2012mid}, does not change our main conclusions. The distribution of GEM and GAMA WISE galaxies in the ($\rm W_1 - W_2$) vs $\rm W_2$ parameter space and the selection criterion is shown in figure \ref{GAMA_WISE}. 

\noindent SED fitting (in principle) should recover each IR-dominated AGN because the SED fits are equally sensitive to the same flux ratios that lead to an IR colour excess. In our analysis, however, as detailed below (\S\,\ref{sec:diag_prospect}), we do not use \textsc{ProSpect}'s AGN mode in the SED fits.
\subsection{The MEx Diagram}\label{sec:diag_mex}
\noindent Traditional nebular-line based diagnostic diagrams suffer from two main issues: the absence of near-infrared (NIR) lines at higher redshifts and the non-zero probability of galaxies near dividing lines being in either of the classes. \cite{2011ApJ...736..104J} introduced the mass-excitation (MEx) diagnostic diagram, a novel probabilistic approach splitting galaxies into sub-categories more confidently than alternative diagnostic diagrams developed to identify AGN at similar redshifts. 

\noindent The line ratios used in the BPT diagram are a combination that probes two important aspects of galaxies: the hardness of their ionising radiation and their gas-phase metallicity. However, at higher redshifts, the lines [NII] and H$\alpha$ become redshifted out of the optical window, excluding their use unless NIR spectra are available. We can leverage other empirical relations and properties of galaxies to mimic this approach to AGN identification. The empirical mass-metallicity relation \citep{tremonti2004origin} of normal galaxies suggests that mass is an indicator of metallicity ([NII]/H$\alpha$). The value of the ratio [NII]/H$\alpha$ is a direct measure of the energy output, which is in turn suggestive of AGN activity (\citealt{2003MNRAS.346.1055K,2006MNRAS.372..961K}). Given these relationships, stellar mass can serve as an excellent proxy for the [NII]/H$\alpha$ ratio at both lower and higher redshifts when it is not available or is challenging to measure accurately. 

\noindent The MEx diagram classifies galaxies into three groups: AGN, composite, and SFG. Figure~\ref{fig:MEx} shows the GAMA and GEM AGN, SFGs and composites identified using the MEx diagram. As with our approach to composite galaxies in the BPT classifier, here we likewise include the MEx composites among AGN. 
\begin{table*}
    \centering
    \begin{tabularx}{\textwidth}{X|X|X|X|X|X}
        \hline
        Parameter & Description & Scale & Units & Range & Prior\\
        \hline
        \verb|mSFR| & peak SFR & Log & $M_\odot$/yr & [-3,6]  & - \\
        \verb|mpeak| & lookback time when the peak SFR occurred & Linear & Gyr & [-2,13.38] & - \\
        \verb|mperiod| & width of the SFH & Log & Gyr & [0.3,3] & - \\
        \verb|mskew| & skewness of the SFH & Linear & - & [-0.5,3] & - \\
        \verb|Zfinal| & final gas-phase metallicity & Log & - & [-4,-1] & - \\
        \verb|alpha_SF_birth| & power law of the radiation heating the birth cloud & Linear & - & [0.01,4] & $\exp\left(\dfrac{-1}{2}\left(\dfrac{\alpha_{birth}-2}{1}\right)^2\right)$\\
        \verb|alpha_SF_screen| & power law of the radiation heating the ISM & Linear & - & [0.01,4] & $\exp\left(\dfrac{-1}{2}\left(\dfrac{\alpha_{screen}-2}{1}\right)^2\right)$\\
        \verb|tau_birth| & optical depth of the birth cloud & Log & - & [-2.5,1] & - \\
        \verb|tau_screen| & optical depth of the ISM screen & Log & - & [-5,1] & - \\
        \hline
    \end{tabularx}
    \caption{The free parameters of the \textsc{ProSpect} fitting code and their characteristics. Different parameters corresponding to modelling the SFH, dust absorption and re-emission, and metallicity are shown in the first column. A short description of each parameter is given in column 2, and the scale (log or linear) adopted while fitting in column 3. Columns 4 and 5 give the units of the free parameters and the range that was chosen to constrain the fitting, respectively. Column 6 shows the prior function used, applied only for the alpha parameters. \textsc{ProSpect} is an interface to the simple stellar population (SSP) models, dust attenuation and re-emission libraries, which then uses the Highlander package (a combination of parameter optimisation and Bayesian inferential techniques) to optimise the free parameters and generate the SED corresponding to the observations.}
    \label{tab:Parameters}
\end{table*}
\subsection{The \textsc{ProSpect} Selection}\label{sec:diag_prospect}
\noindent \textsc{ProSpect} \citep{2020MNRAS.495..905R,2020MNRAS.498.5581B,2022MNRAS.509.4940T} is a  generative galaxy SED fitting package. It is equipped with the stellar population libraries BC03 \citep{2003MNRAS.344.1000B} and EMILES \citep{2016MNRAS.463.3409V}, both using a fixed Chabrier initial mass function (IMF) \citep{Chabrier_2003}. The package incorporates the \cite{2000ApJ...539..718C} dust attenuation model, as well as the dust re-emission library of FIR templates presented by \cite{2014ApJ...784...83D}. 

\noindent Initially, \textsc{ProSpect} allowed fitting of SEDs only extending to the FIR region. In a recent extension by \cite{2023MNRAS.522.6354T}, the package now supports fitting up to radio wavelengths. The radio flux densities in these models are estimated from the FIR fluxes through the link between FIR-derived star formation rates (SFRs) and thermal and non-thermal radio emission. They adopt power-laws with $\alpha=-0.1$ and $\alpha=-0.7$ (where $S_\nu\propto\nu^\alpha$) for these free-free and synchrotron components respectively \citep{2023MNRAS.522.6354T,2015AJ....149...32M}. This forward-modelled radio flux solely originates from star formation processes. The simulated radio flux is then compared with the EMU radio data to distinguish between AGN and SFGs. Any excess radio emission above a defined threshold can be considered as an indication of AGN activity.
\begin{figure}[h!]
    \centering
    \includegraphics[width=0.9\linewidth]{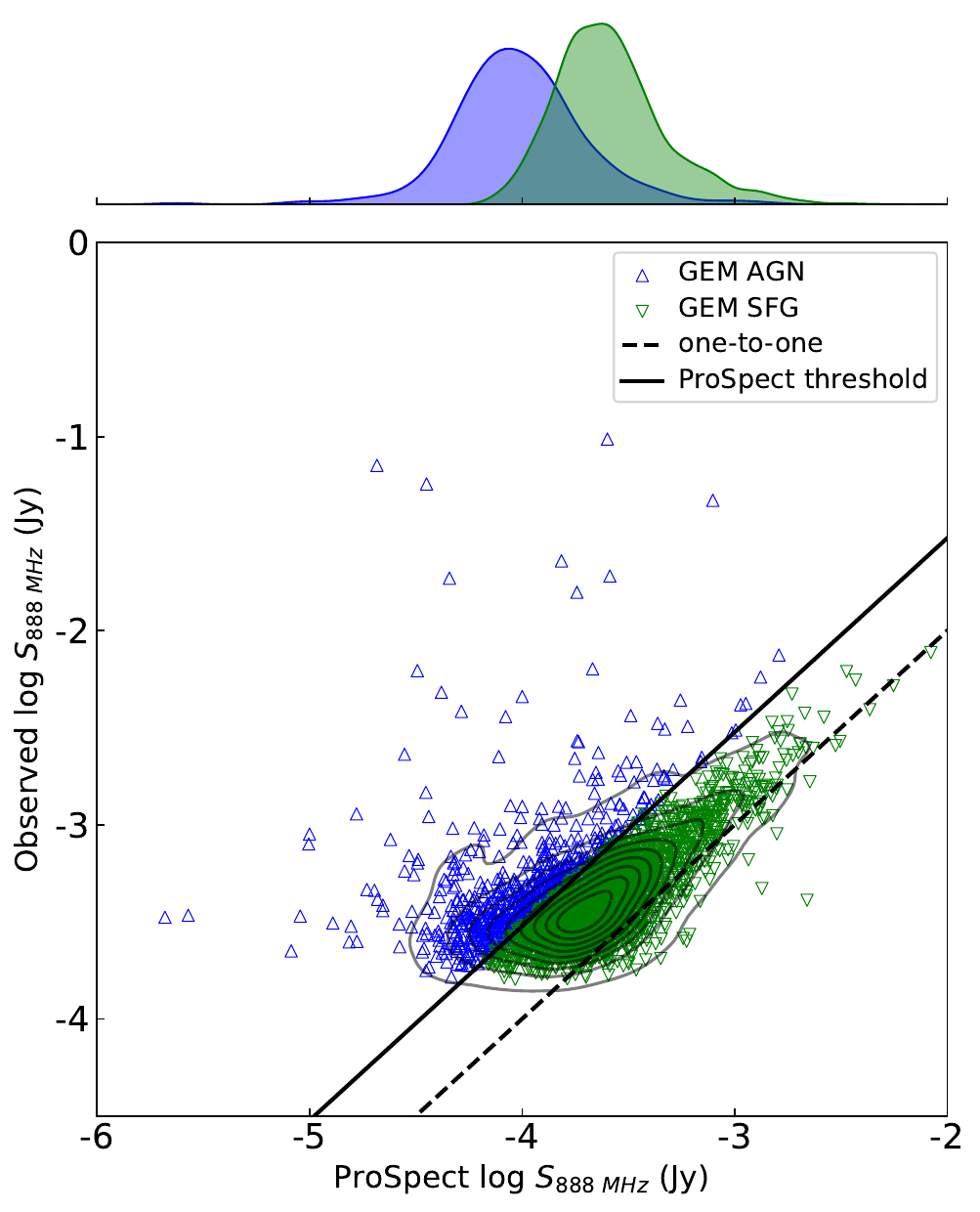}
    \caption{\textsc{ProSpect} AGN selection. The new SED-based AGN diagnostic scheme showing the relation between observed EMU radio flux and the predicted radio flux from SF processes, both at 888 MHz. The blue  triangles are the GEM \textsc{ProSpect} AGN and the green  triangles are the GEM \textsc{ProSpect} SFGs. The demarcation (solid line) between the two corresponds to the observed radio flux being at least 3 times the predicted radio flux from SF. The dashed line shows the one-to-one relation.}
    \label{fig:ProSpect}
\end{figure}

\noindent To model the SFH, we use the \verb|massfunc_snorm_trunc| parameterisation, where \verb|mpeak|, \verb|mSFR|, \verb|mperiod|, and \verb|mskew| are set as free parameters. For the metallicity history,  we use the \verb|Zfunc_massmap_lin| keeping \verb|Z_final| as a free parameter. These parameters play crucial roles in determining the stellar mass and star formation rate of the galaxy at different epochs, taking into account any variations in the SFH and metallicity over time. 

\noindent For modelling dust, we include four additional free parameters: \verb|alpha_SF_birth|, \verb|alpha_SF_screen|, \verb|tau_birth|, and \verb|tau_screen|. The \verb|alpha_SF_birth| and \verb|alpha_SF_screen| parameters represent the intensity of the dust radiation field for the birth and ISM clouds, respectively, whereas \verb|tau_birth| and \verb|tau_screen| parameters are measures of dust opacity for the same. To facilitate the fitting process, a prior function is included for the \verb|alpha| parameters \citep{2020MNRAS.498.5581B}. This prior function helps guide the fitting process and enables \textsc{ProSpect} to better model the SED and derive more accurate physical properties faster. Table \ref{tab:Parameters} provides a summary of the different parameters used in our SED fitting process and their characteristics.

\noindent The FUV-FIR photometry gives the flux densities in twenty different bands. We note here that some objects do not have photometry in certain bands. In such cases, the SED was fitted using only the available photometry. We use floor values to supplement the uncertainties associated with each band to ensure reliable fits even when reported uncertainties are too optimistic (FUV = 0.2, NUV = 0.3, u = 0.2, g = 0.2, r = 0.2, i = 0.2, Z = 0.2, Y = 0.2, J = 0.2, H = 0.2, Ks = 0.2, $\rm W_1=0.05,W_2=0.14,W_3=0.145,W_4=0.16$, P100 = 0.001, P160 = 0.001, S250 = 0.001, S350 = 0.001, and S500 = 0.001) and the final fitting uncertainties are calculated as described in \cite{2020MNRAS.498.5581B}.

\noindent In the fitting process, we use the Highlander\endnote{\url{https://github.com/asgr/Highlander}} package to model the SED of a galaxy. We implement this in a Bayesian manner using the Componentwise Hit and Run Metropolis (CHARM) algorithm, which is a Markov Chain Monte Carlo (MCMC) method provided by the \verb|LaplacesDemon| (LD) package in \verb|R| \citep{LD}. For each galaxy, MCMC is run in two iterations, with each iteration optimising the parameters in 1000 steps and then feeding the optimised parameters into LD (500 steps in the first iteration and 1000 steps in the second iteration).
\begin{table}
    \centering
    \begin{tabular}{c|c|c|c|c}
    \hline
        Identifier & BPT & WISE & \textsc{Pro}\textsc{Spect} & MEx \\
        \hline
        0000 & SFG & SFG & SFG & SFG \\
        0001 & SFG & SFG & SFG & AGN \\
        0010 & SFG & SFG & AGN & SFG \\
        0011 & SFG & SFG & AGN & AGN \\
        0100 & SFG & AGN & SFG & SFG \\
        0101 & SFG & AGN & SFG & AGN \\
        0110 & SFG & AGN & AGN & SFG \\
        0111 & SFG & AGN & AGN & AGN \\
        1000 & AGN & SFG & SFG & SFG \\
        1001 & AGN & SFG & SFG & AGN \\
        1010 & AGN & SFG & AGN & SFG \\
        1011 & AGN & SFG & AGN & AGN \\
        1100 & AGN & AGN & SFG & SFG \\
        1101 & AGN & AGN & SFG & AGN \\
        1110 & AGN & AGN & AGN & SFG \\
        1111 & AGN & AGN & AGN & AGN \\
        \hline
    \end{tabular}
    \caption{The binary classification scheme}
    \label{tab:bin}
\end{table}

\noindent To estimate the radio emission from the SED, we enable the \verb|addradio_SF=TRUE| option. The underlying concept is to model the SED of the galaxy assuming no AGN contribution, which results in an estimate of the radio emission purely from star formation processes. An example SED of a galaxy with the contributing factors (stellar emission, dust absorption, and re-emission) is shown in figure \ref{fig:SED}. To distinguish between AGN and SFGs, we employ a threshold for the ratio of the observed EMU radio data to the radio flux estimate obtained from the SED fitting without AGN. 

\noindent For this analysis, and to be conservative, we choose a threshold of 3. This choice is somewhat arbitrary and could be lower or higher depending on how the diagnostic is to be used. We did check the impact on our AGN selection of varying the threshold value, and, as expected, lowering the threshold increases the sample of potential \textsc{ProSpect} AGN, but also increases the number classified as SF galaxies by the other classifiers, implying greater contamination. For the current process, we are aiming to demonstrate that this is a reliable AGN diagnostic, and accordingly, we choose a relatively high threshold so as not to bias our result through contamination by star formation from fits with large uncertainties. 

\noindent The FIR-radio correlation \citep{condon1992radio} exhibits a mass dependency \citep{2018MNRAS.475.3010G,2021A&A...648A...6S,2021A&A...647A.123D}. Low mass galaxies exhibit a higher value of $\rm q_{IR}$ \citep[$L_{IR}/L_{1.4~GHz}$;][]{2021A&A...647A.123D} than the high mass galaxies \citep[see figure 14 of ][]{2021A&A...647A.123D}. This can be attributed to the escape of cosmic ray electrons (CRe) from low mass galaxies before they radiate away their energies. The consequence of this is a reduction in the radio luminosity of the low mass galaxies from that predicted by the linear FIR-radio correlation. This may result in some low mass galaxies which would be AGN but not be selected by the \textsc{ProSpect} method. The FIR-radio correlation is tight with a scatter of only 0.26 \citep[e.g.][]{2001ApJ...554..803Y}, corresponding to a factor of 1.9. With the \textsc{ProSpect} selection threshold of 3, this would not result in any SFG being misclassified as AGN. If anything, this conservative selection might exclude some AGN.
\begin{figure}[h!]
    \centering
    \includegraphics[width=0.95\linewidth]{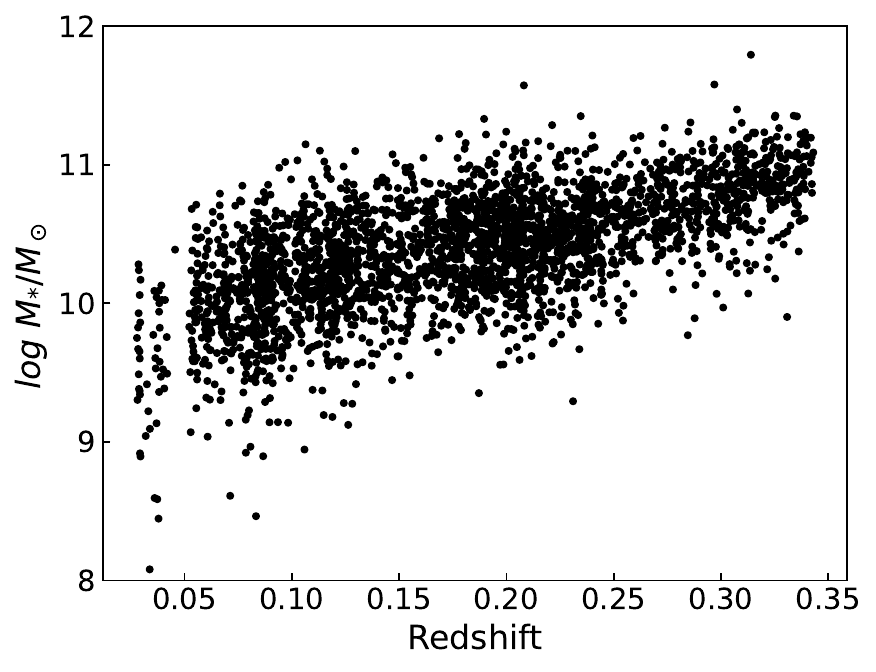}
    \caption{Variation of stellar masses of the GEM galaxies with redshift. The requirement of reliable spectral lines for the emission line diagnostics, especially the BPT diagram, confines the redshift of the GEM sample to a maximum of $z=0.34$. At higher redshifts,  the [NII] and H$\alpha$ lines are redshifted out of the optical window. The typical trend seen in flux-limited samples is visible, with an apparent trend of increasing mass with redshift. This is a consequence of the flux limit excluding low mass galaxies at higher redshift, while at low redshift, the sampled volume is not large enough to capture rare massive galaxies.}
    \label{fig:mass_z}
\end{figure}
\begin{figure*}[ht!]
    \centering
    \includegraphics[width=0.9\linewidth]{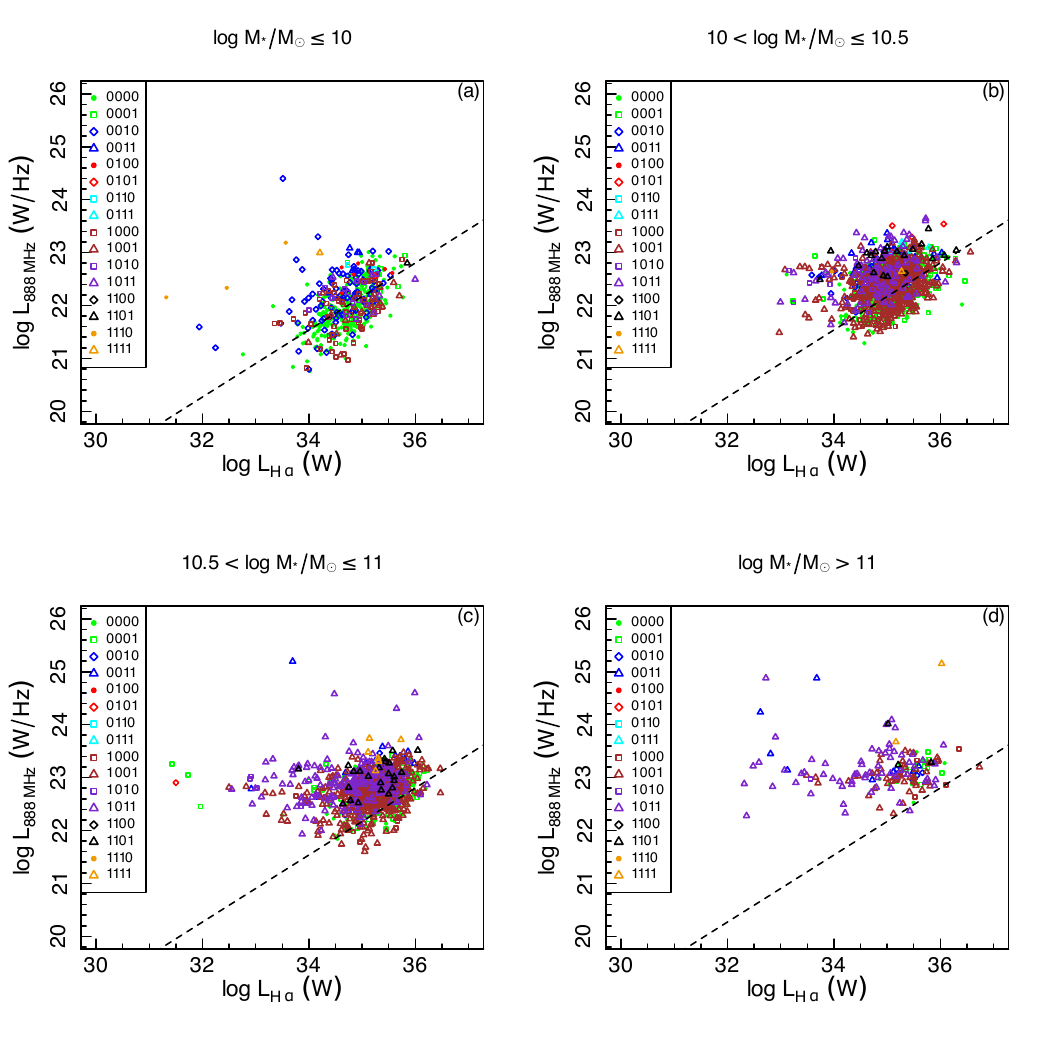}
    \caption{888 MHz luminosity as a function of H$\alpha$ luminosity for four different stellar mass bins. The colours of the data points are selected to aid in distinguishing the 16 classes. The emission-line detected AGN are dominant in the three highest mass bins, whereas a large number of AGN in the lowest mass bin are identified by \textsc{ProSpect}. The dashed line is a linear fit to the population in the lowest mass bin and is shown for reference in other mass bins. This emphasises the upward shift of galaxies in luminosity as mass increases, along with an increased scatter in the radio excesses.}
    \label{fig:H-radio}
\end{figure*}

\noindent Figure \ref{fig:ProSpect} presents a comparison between the observed and the predicted radio fluxes based on SED fitting. This is effectively an AGN selection criterion based on a  radio emission threshold. Galaxies that fall above this threshold are identified as AGN, while those below it are classified as SFGs. An apparent asymmetry is evident in the distribution of AGN and SFGs around the one-to-one line. This is associated with our sample selection, as we require the objects to be radio-detected. SFGs that would otherwise contribute to a symmetrical distribution may fall below the radio sensitivity limit of the survey and, hence, are not represented here.

\noindent Given the challenge of tracking which classification each galaxy has from each diagnostic, we introduce a binary classification scheme. Galaxies are labelled \verb|0| and \verb|1|, meaning they are SFG or AGN dominated, respectively, in a particular diagnostic and the scheme is detailed in table \ref{tab:bin}.
\begin{figure*}[h!]
    \centering
    \includegraphics[width=0.9\linewidth]{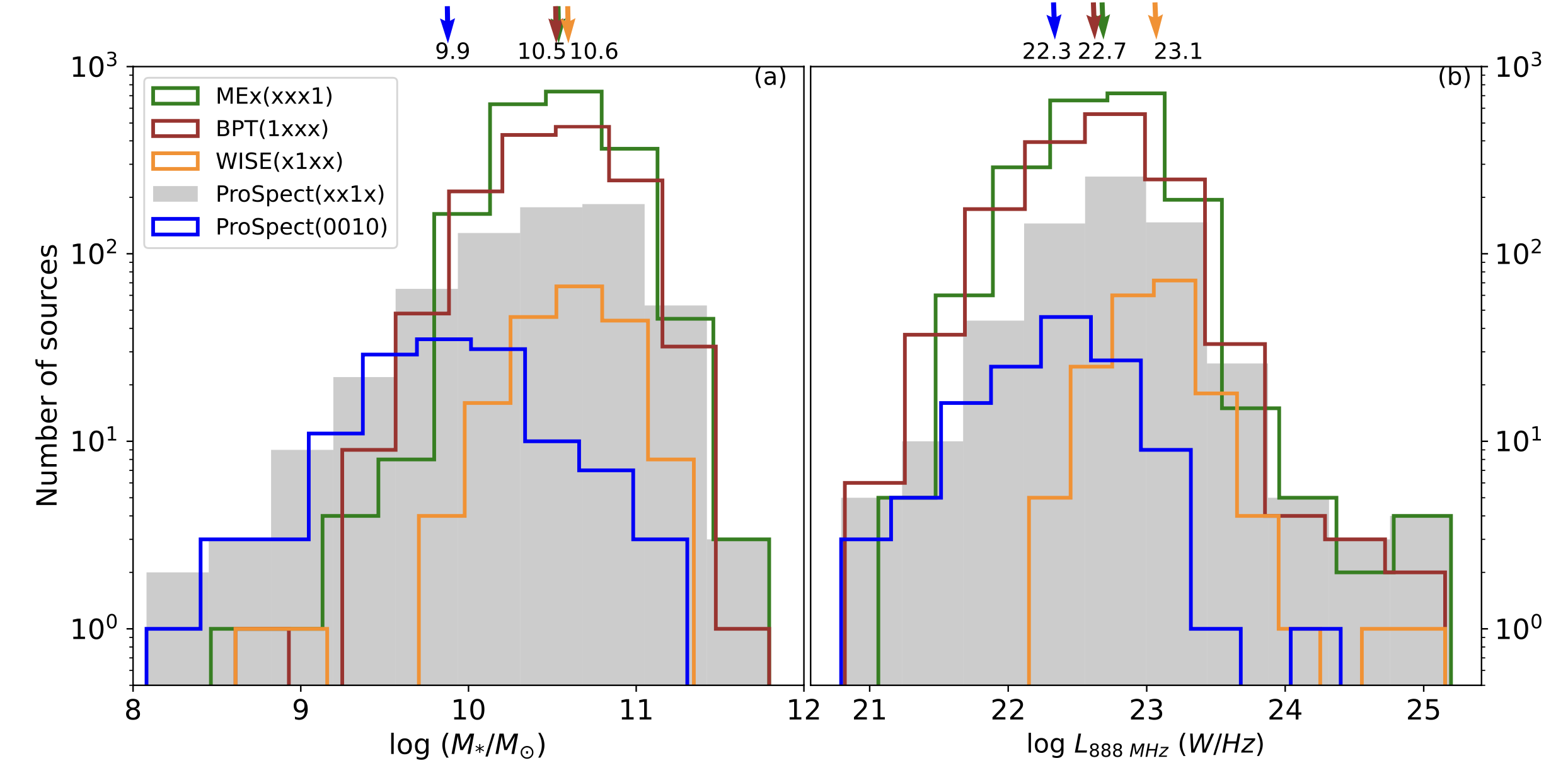}
    \caption{Panel (a): Distribution of host galaxy stellar mass for AGN identified by different classifiers. AGN detected by narrow-line diagnostics are shown in green and brown, and WISE detected AGN in orange. The grey filled distribution shows the AGN identified by multiple classifiers including \textsc{ProSpect} and the \textsc{ProSpect} alone AGN are represented by the blue histogram. The arrows coloured according to the histograms show the mean (the medians are almost identical to the mean) of these distributions in the respective colours (BPT and MEx have the same mean values). Panel (b): Distribution of host galaxy radio luminosity for AGN identified by different classifiers. The colours of the distributions and the arrows are as in panel (a). The distribution of ProSpect AGN clearly favours the low mass, low radio luminosity systems, while the distributions of AGN detected by multiple classifiers favour higher host galaxy mass and radio luminosities.}
    \label{fig:mass_lum_hist}
\end{figure*}
\section{Results}\label{sec:results}
\noindent As \textsc{ProSpect} identification is based on radio excess, we can have a high level of confidence in the AGN detected by this method. In this section, we explore and compare various physical properties of the AGN identified by the classifiers. We investigate whether \textsc{ProSpect} provides unique information beyond what other methods offer. The analyses in the upcoming subsections use the binary identifier scheme introduced in section \ref{sec:diagnostics}. Table \ref{tab:bin} is a guide to the identification scheme.
\subsection{Luminosities: H$\alpha$ - Radio}\label{sec:disc_ha}
\noindent Figure \ref{fig:mass_z} shows how the host stellar masses of our sample are distributed over redshift. The deficit of low mass hosts at higher redshifts is a consequence of the flux limit of the survey whereas that of massive galaxies in the local universe is due to the small volume sampled in that redshift range.
\noindent Figure \ref{fig:H-radio} shows the relation between H$\alpha$ and 888 MHz luminosities in four different stellar mass bins. We choose to represent the data by looking at the relationship between these luminosities, as both luminosities are sensitive to star formation rate. Excesses above a one-to-one correlation in this diagram would imply an excess radio emission that could be attributed to a radio AGN, although low (radio) luminosity AGN will not necessarily show such an excess. 

\noindent We chose the division in mass to have a roughly similar number of objects in the middle two bins, which contain the most galaxies, while still separately sampling the highest and lowest mass ranges. Table \ref{tab:AGN_numbers} presents the number of galaxies in these mass bins arising from the different combinations of AGN indicators. The colours chosen in the key have been carefully selected to help guide the eye in interpreting this very busy figure (and those to follow). In broad terms, and neglecting the MEx classifier (for simplicity in the first instance), the green colours are galaxies classified as star forming, blue are those classified as AGN by \textsc{ProSpect} (or \textsc{ProSpect} and MEx), and brown are those classified as AGN by BPT (or BPT and MEx). The red symbols (WISE, or WISE and MEx) are very few in number, and the remaining colours, purple, black, and yellow, refer to galaxies classified as AGN by multiple indicators.

\noindent Most AGN, especially those identified by BPT and MEx, populate the two middle mass bins, and it is noticeable that very few WISE and relatively less MEx-identified AGN exist in the lowest mass bin. The lowest masses, however, do contain a significant number of AGN, and these are primarily identified through the \textsc{ProSpect} and BPT classifiers. This is reflected in the numbers in table \ref{tab:AGN_numbers}. From this initial interpretation, it appears that \textsc{ProSpect} is effective at identifying AGN in galaxies of low mass, some of which demonstrate low radio and H$\alpha$ luminosities. The distribution of AGN by indicator as a function of host galaxy stellar mass and as a function of radio luminosity is shown in figure \ref{fig:mass_lum_hist}, which reinforces this result.

\noindent In the lowest mass bin (figure \ref{fig:H-radio}), there is a trend between the radio and H$\alpha$ luminosities for most galaxies (indicated by the dashed line). Those AGN identified by the BPT follow this trend, but those identified by \textsc{ProSpect} fall above or on the upper edge of this trend, as expected from radio excess systems. In the higher mass bins, the populations move to higher luminosities and have an increased scatter of radio excess objects. This is a consequence of the variation of host galaxy mass with redshift. 
\begin{table*}
    \centering
    \begin{tabularx}{\textwidth}{m{2.5cm}|X|X|X|X|X|X|m{2cm}}
        \hline
        mass\_bin&AGN/Total&BPT AGN&MEx AGN&\textsc{ProSpect} AGN&BPT alone&MEx alone&\textsc{ProSpect} alone\\
        & &(\verb|1xxx|)&(\verb|xxx1|)&(\verb|xx1x|)&(\verb|1000|)&(\verb|0001|)&(\verb|0010|)\\
        \hline
        log$(M_\textbf{*}/M_\odot)\leq$10&206/481&95&57&113&51&17&81 \\
        10<log$(M_\textbf{*}/M_\odot)\leq$10.5&970/1269&583&857&217&56&261&38 \\
        10.5<log$(M_\textbf{*}/M_\odot)\leq$11&948/1051&659&903&240&21&201&11 \\
        log$(M_\textbf{*}/M_\odot)$>11&148/155&122&137&77&6&11&3
    \end{tabularx}
    \caption{The numbers of AGN in different mass bins identified by different classifiers. The first column shows the mass bins and the total AGN numbers in each mass bin are shown in the second column. Here, for clarity in the display, we neglect the WISE AGN due to their small numbers. Columns three to five list the AGN detected by BPT, MEx, and \textsc{ProSpect} regardless of the identification by other classifiers. Columns six to eight show the AGN detected by BPT, MEx, and \textsc{ProSpect} alone.}
    \label{tab:AGN_numbers}
\end{table*}

\noindent It is worth noting that even in the highest mass bin (log $\rm M_\textbf{*}/M_\odot$>11) we still find galaxies classified as star forming only by all four classifiers. We have visually inspected the spectra of all 13 of these potential SFGs and confirmed that 6 of them are genuine star formers with no apparent AGN signature in the indicators we have utilised. The rest were assigned manually to the BPT AGN class. These galaxies lie in the upper range of our redshift space, with z>0.1 for all massive star formers.

\noindent A close look at figures \ref{fig:H-radio}, \ref{fig:mass_lum_hist}, and table \ref{tab:AGN_numbers} clearly demonstrates that the low mass end of the AGN population in our GEM sample is roughly equally populated by BPT and \textsc{ProSpect} AGN. With the mean mass of the \textsc{ProSpect} AGN population being $\rm 10^{9.9}~M_\odot$, \textsc{ProSpect} is efficient at detecting low mass AGN systems which would not have been identified with narrow line and mid-IR diagnostics alone. Though BPT, in combination with other methods, detects a fairly large number of AGN in the lowest mass bin, the \textsc{ProSpect} method detects the largest number of AGN here. 
\begin{figure}[h!]
    \centering
    \includegraphics[width=0.95\linewidth]{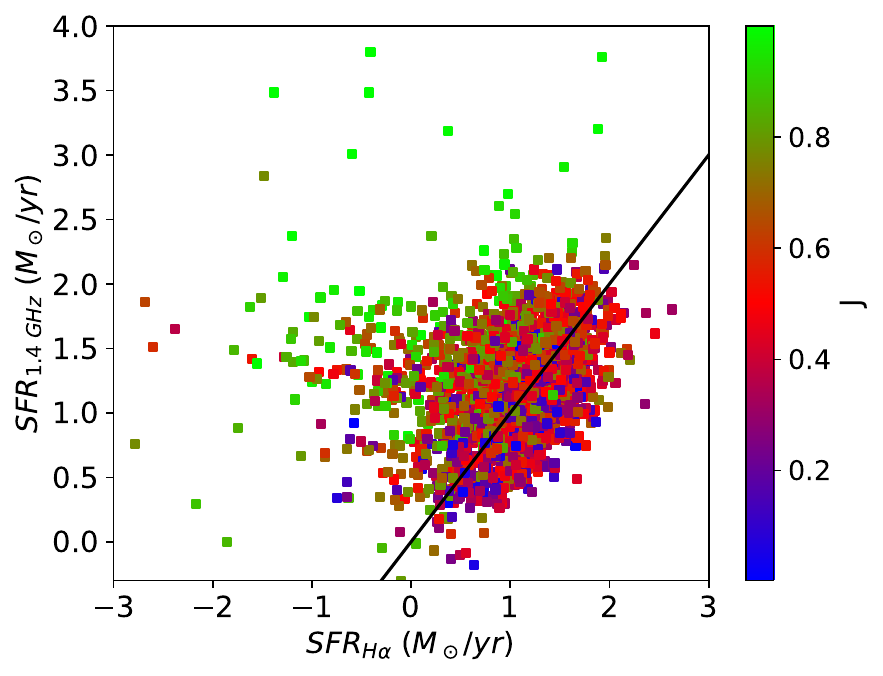}
    \caption{1.4 GHz radio SFR as a function of H$\alpha$ SFR. The 1.4 GHz radio SFR are calculated after converting the 888 MHz radio luminosities to 1.4 GHz. The solid black line represents the one-to-one correlation between both SFRs. The points are colour-coded according to J. The sources with higher values of J ($\geq 0.67$) exhibit excess radio emission.}
    \label{fig:j_sfr}
\end{figure}
\subsection{Radio luminosity fractional AGN contribution}\label{sec:disc_agn_frac}
\noindent We define the parameter $J$ as the ratio of excess radio luminosity compared to that predicted by \textsc{ProSpect} ($\rm L_{Obs}-L_{\textsc{ProSpect}}$) to the observed radio luminosity:
\begin{equation*}
    J = \frac{L_{Obs}-L_{\textsc{ProSpect}}}{L_{Obs}}.
    \label{eq:4.1}
\end{equation*}
\noindent The quantity $J$ is the ratio of the AGN's contribution to the total radio luminosity of the galaxy. It answers the question, ``What fraction of a galaxy's radio emission is from AGN as opposed to star formation?'' 
Since there are galaxies that lie below the one-to-one line in Figure~\ref{fig:ProSpect}, there will be systems with a negative value of $J$. These by definition are classified as star forming by the ProSpect diagnostic, but could still be classified as AGN by other methods.

\noindent We show how $J$ varies in Figure~\ref{fig:j_sfr}, which presents radio SFR as a function of H$\alpha$ SFR, colour-coded by $J$. The radio SFR here is calculated after converting our 888\,MHz luminosities to 1.4\,GHz, assuming a radio spectral index of $\alpha=-0.7$, and the SFR calibration of \citet{2003ApJ...599..971H}. This confirms that sources with an excess of radio emission over the otherwise broad correlation have higher values of $J$, as expected.

\begin{figure*}[ht!]
    \centering
    \includegraphics[width=0.9\linewidth]{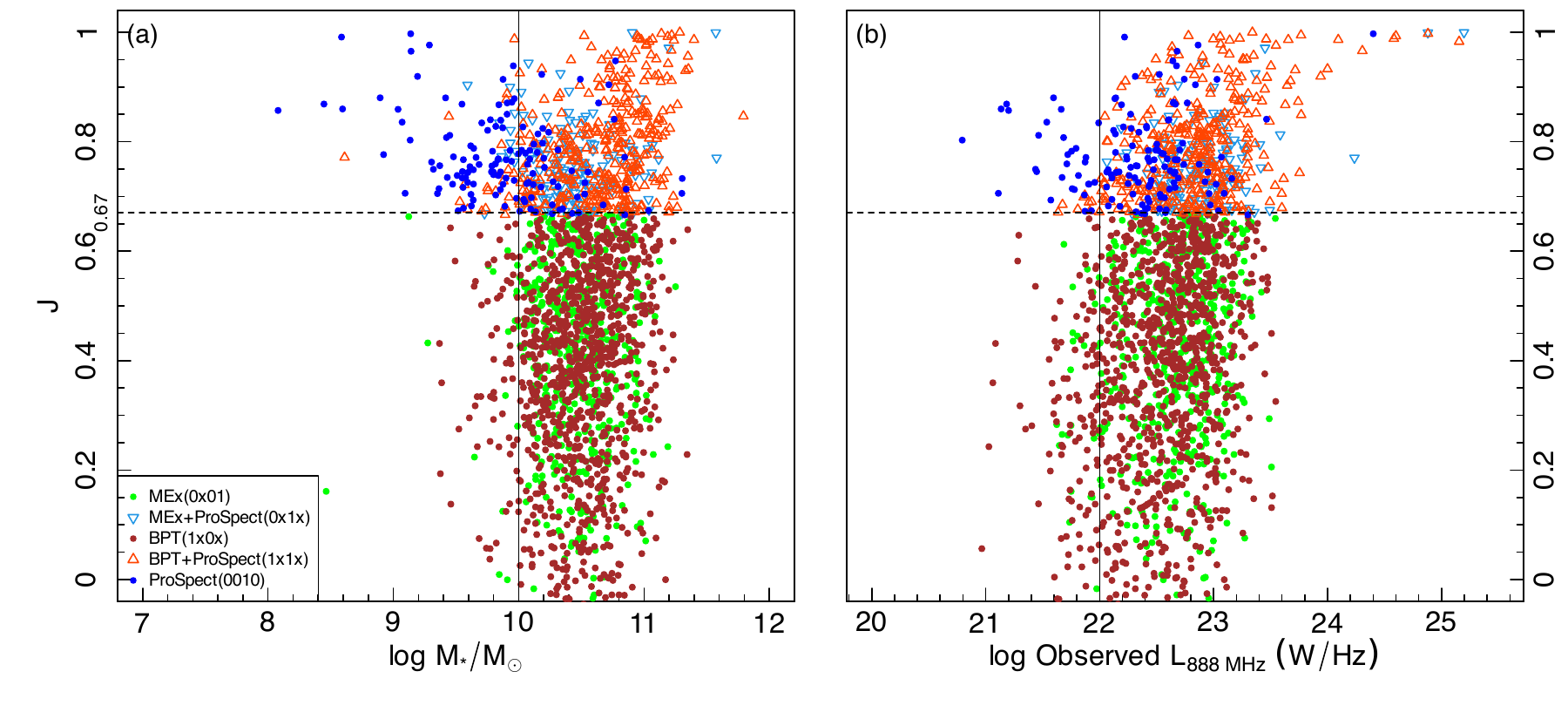}
    \caption{The radio luminosity fractional AGN contribution ($J$) as a function of (a)~stellar mass and (b)~radio luminosity. For conciseness in the key, we use an ``x'' to avoid repetition of both 1 and 0 where classifications are consolidated. So, e.g., 0x01 means both 0001 and 0101. That is, these are objects classified as AGN by MEx and SF by ProSpect and BPT, independent of the WISE classification. To be clear we spell out here what each colour represents in terms of explicitly which binary classifications are jointly included. The colour scheme is identical for both panels. Green (MEx AGN: 0001, 0101); Cyan (MEx+\textsc{ProSpect} AGN: 0011, 0110, 0111); Brown (BPT AGN: 1000, 1001, 1100, 1101); Orange (BPT+\textsc{ProSpect} AGN: 1010, 1011, 1110, 1111); Blue (\textsc{ProSpect} AGN only: 0010). Any datapoint below $J=0.67$ (the horizontal dashed line) will, by construction, not be identified by \textsc{ProSpect}. Consequently, the green and brown dots are AGN identified by all classifiers except ProSpect. Above the ProSpect threshold, the blue dots are AGN detected by \textsc{ProSpect} alone, whereas the orange and cyan triangles are AGN identified by multiple identifiers, including \textsc{ProSpect}. The figure clearly demonstrates that ProSpect is sensitive to AGN  hosted by low mass galaxies with low radio luminosities. The solid line represents the low mass threshold in (a) and the low radio luminosity threshold in (b).}
    \label{fig:AGN_frac}
\end{figure*}

\noindent Figure \ref{fig:AGN_frac} presents this parameter as a function of mass and radio luminosity. Here we only show positive values of $J$. The \textsc{ProSpect} detected sources, by definition, lie at $J>0.67$ because of our threshold value of 3 (defined in \S\, \ref{sec:diag_prospect}), which we call the ``\textsc{ProSpect} threshold''. The transition to different AGN diagnostics being dominant above log ($\rm M_\textbf{*}/M_\odot) = 10$, reinforces the preferential selection of AGN from low mass hosts by \textsc{ProSpect}.

\noindent Above the \textsc{ProSpect} threshold, AGN show a clear demarcation of diagnostics in the region of parameter space they occupy in figure \ref{fig:AGN_frac}. \textsc{ProSpect} AGN, hosted by low mass galaxies, extend towards the low radio luminosity end to a degree not seen with other indicators. Fully one-fifth (22\%) of the \textsc{ProSpect} AGN lie at $L_{888~\rm MHz} < 10^{22}~\rm W/Hz$. This compares to a fraction of one-tenth (10\%) for the BPT AGN. The population exhibits a mixing at mid-radio luminosity giving way to AGN detected by multiple diagnostics at high radio luminosity end. Below the \textsc{ProSpect} threshold, AGN lie predominantly at higher radio luminosities, with very few extending to the faintest levels. This is true also for those non-ProSpect AGN with negative $J$ values.
\section{Discussion}\label{sec:discussion}
\subsection{Comparison with a low mass AGN sample}\label{sec:disc_low_mass}
\noindent \cite{2022ApJ...937....7S} identified 388 low mass AGN hosts in the GAMA DR4 equatorial and  G23 regions consistent with our definition of low mass ($\rm \log M_\textbf{*}/M_\odot\leq10$). Of this sample, 53 lie in G23. Their search was primarily using narrow-line diagnostic techniques, including the BPT diagram. A cross-match between the 53 low mass AGN hosts and the EMU early science data with a 5" radius resulted in 11 common sources. Thus the radio selection is the primary reason for the large difference in their sample and ours. A cross-match between these 11 low mass AGN hosts and our GEM sample resulted in 9 common sources whose details are shown in table \ref{tab:low_mass}. One of these is classified by all four classifiers as an SFG in our sample. The two missing sources in the GEM sample are removed by the LP selection described in \S\,\ref{sec:data_final}.

\noindent While \textsc{ProSpect} identifies 2 out of the 8 remaining AGN in common, none of the AGN detected by \textsc{ProSpect} alone in our GEM sample are present in their low mass AGN sample. This is consistent with the above results demonstrating that narrow-line techniques are not complete in identifying AGN in low mass galaxies. While narrow-line diagnostic tools can identify bona fide AGN, such identification of AGN from low mass hosts is likely to be more incomplete than at high masses (\citealt{2013ApJ...775..116R}), emphasising the need for the use of a suite of diagnostics to optimise the completeness of AGN samples.

\noindent It should be noted here that this comparison does not check the completeness of the AGN population detected by \textsc{ProSpect}. Our intention is to compare the low mass AGN samples and check whether the \textsc{ProSpect} AGN are unique, which seems to be the case thereby reinforcing the importance of multiwavelength analysis.
\subsection{Deciphering the 0010s}\label{sec:disc_0010}
\noindent We have established that \textsc{ProSpect} is efficient at picking up AGN from low mass systems, which are potentially weak radio emitters. Radio AGN are classified as weak-line radio galaxies (WLRGs) or strong-line radio galaxies (SLRGs) based on their optical emission line intensities \citep{1998MNRAS.298.1035T}. With respect to their properties, WLRGs and SLRGs are almost similar to LERGs and HERGs, respectively \citep{2016A&ARv..24...10T}. 
\textsc{ProSpect} clearly identifies both strong-line and weak-line species. It detects HERGs as it is sensitive to some of the higher mass, higher radio luminosity AGN that are also identified by the emission line techniques. It also identifies low mass and low radio luminosity AGN that are most likely to be LERGs, as they are not identified by the emission line techniques. 
\begin{table}[h!]
    \centering
    \begin{tabular}{c|c|c|c}
    \hline
         CATAID&Identifier&log($\rm M_\textbf{*}/M_\odot)$&log $\rm L_{888~MHz}$ \\
         \hline
         5154526&0000&9.8&21.8\\
         5364072&1000&9.7&21.7\\
         5280475&1000&9.8&21.4\\
         5241739&1000&9.8&22.9\\
         5266552&1000&9.9&22.6\\
         5197149&1000&9.9&22.5\\
         5240995&1001&9.9&21.8\\
         5368772&1010&9.7&22.4\\
         5230865&1011&9.9&22.5\\
         \hline
    \end{tabular}
    \caption{The cross-match between the GEM and low mass galaxy samples. There are 8 galaxies common to both datasets, with 7 identified by multiple classifiers and 1 SFG. The absence of AGN unique to \textsc{ProSpect} means that the SED-based AGN hosted by low mass galaxies in our GEM sample is a novel set. We can be confident that the \textsc{ProSpect} AGN are true AGN since the detection is rooted in radio excess.}
    \label{tab:low_mass}
\end{table}

\begin{figure}[h!]
    \centering
    \includegraphics[width=0.98\linewidth]{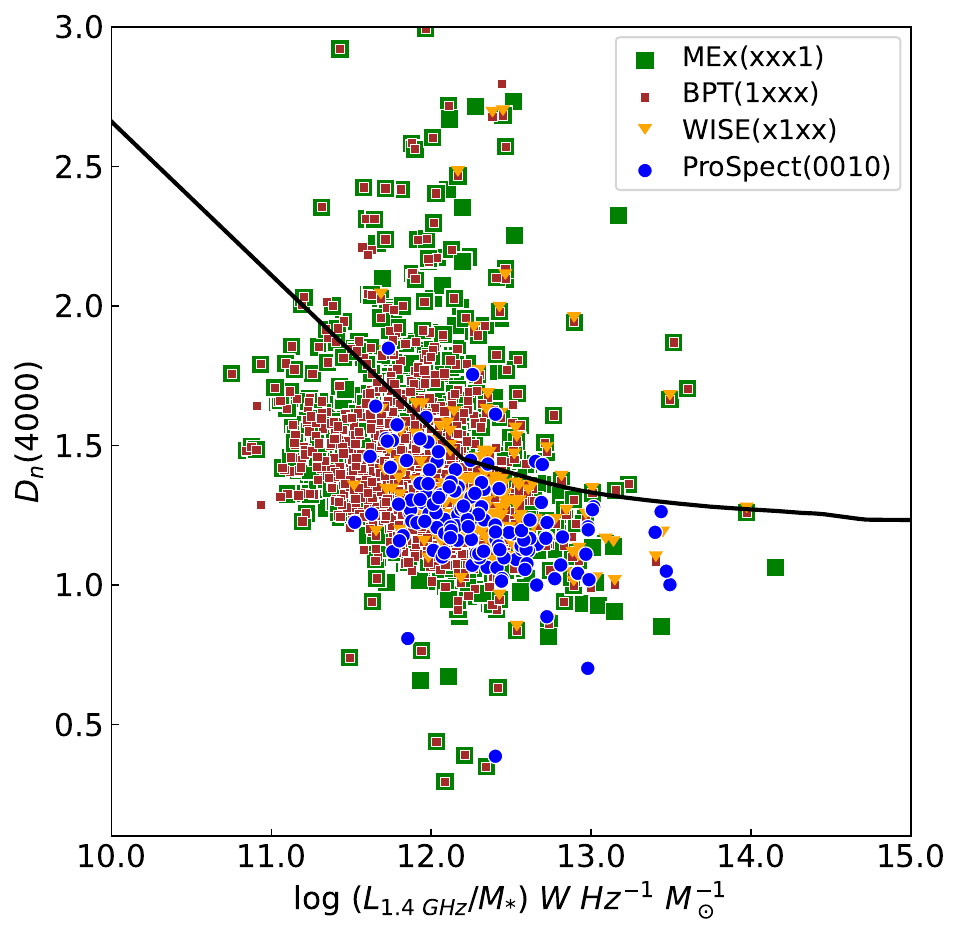}
    \caption{$\rm D_n(4000)$ as a function of 1.4 GHz radio luminosity per galaxy stellar mass. The blue dots are the AGN identified by \textsc{ProSpect} alone, most of them which lie below the selection line. Brown and green squares represent the BPT and MEx AGN, respectively. The orange triangles are WISE AGN. The solid line delineates the optically active radio AGN (above) from the optically weak radio AGN (below). It corresponds to the 3 Gyr exponentially decaying SF track as prescribed in \cite{2012MNRAS.421.1569B}. By definition, SFGs with radio emission occupy the region below the demarcation line. The presence of \textsc{ProSpect} alone AGN in this region indicate that these are low-excitation species with a radio excess.}  
    \label{Dn4000}
\end{figure}

\noindent Figure \ref{Dn4000} shows the 1.4 GHz radio luminosity over galaxy stellar mass as a function of the $\rm D_n(4000)$ index. This method was introduced by \cite{2005MNRAS.362....9B}, and they derived a demarcation between radio-loud AGN and SF galaxies using BC03 SSP models. The bulk of the AGN identified by \textsc{ProSpect} alone (blue) lie in the region of this diagram identified by \cite{2012MNRAS.421.1569B} as LERGs. Accordingly, those AGN identified by \textsc{ProSpect} alone are most likely to be AGN in this low-excitation state.

\noindent We explain this result in terms of proportionally lower mass black holes in low mass galaxies as opposed to massive systems. Black holes are assumed to exist in the centres of almost all galaxies. The influence that these central engines have on their host galaxies can significantly affect their evolution, which in turn can have a feedback effect on the black holes. The proportion of black hole mass has been found to vary as a function of the mass of the host galaxy \citep[see, e.g., figure 3 from][]{2023MNRAS.520.1975G} such that lower mass galaxies host proportionally smaller mass black holes when compared to massive galaxies. Quantitatively, while a $\rm M_*=10^{12}~M_\odot$ galaxy typically hosts a $ \rm 10^{10}~M_\odot$ black hole, a $\rm M_*=10^{9}~M_\odot$ galaxy is found to typically host a black hole of mass of only $\rm 10^{5}~M_\odot$. 
\begin{figure}[h!]
    \centering
    \includegraphics[width=1\linewidth]{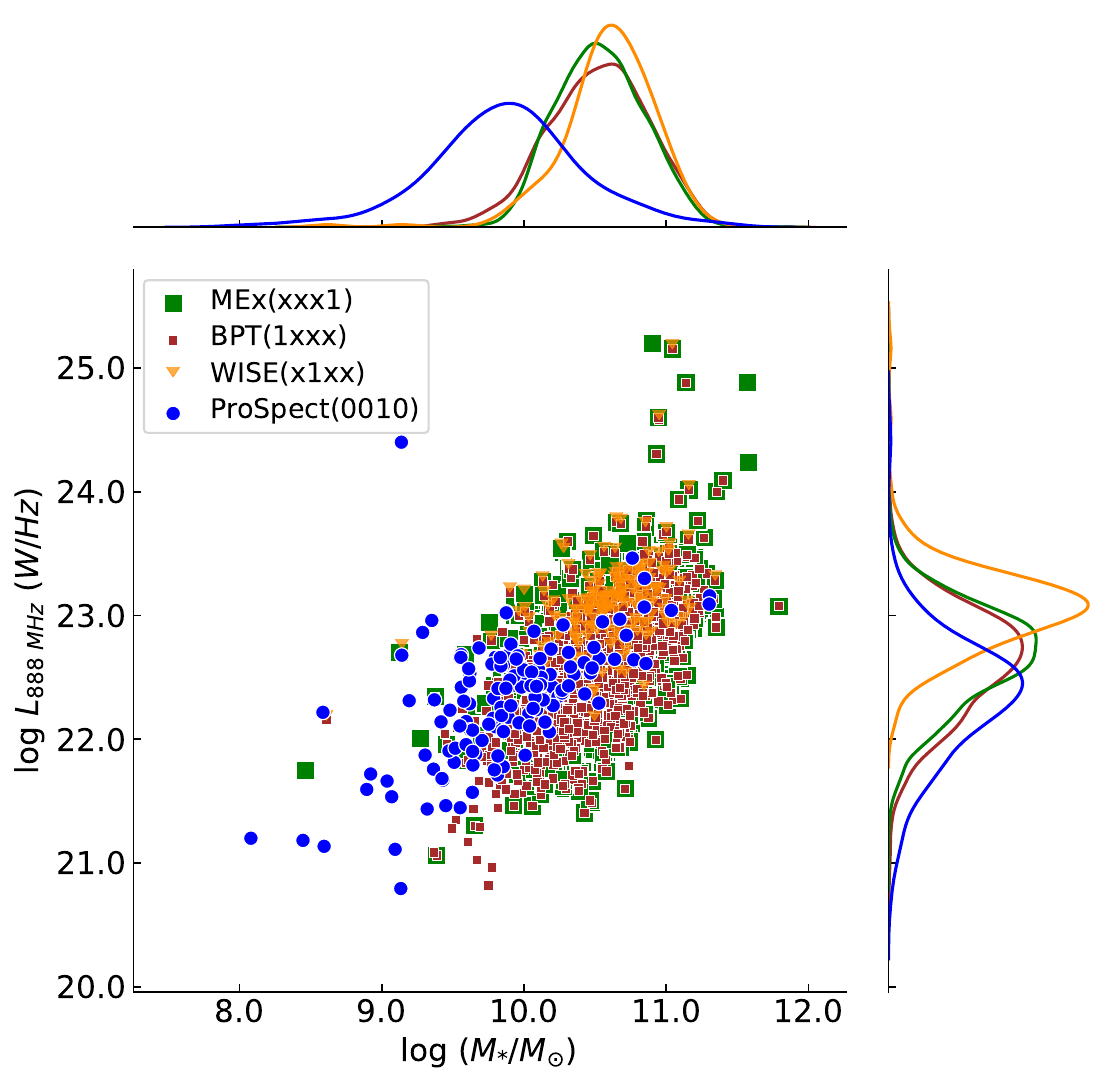}
    \caption{The variation of observed 888 MHz radio luminosities with stellar mass. The blue dots are the AGN identified by \textsc{ProSpect} alone, which preferentially correspond to lower host stellar mass and radio luminosity. Other data points are AGN identified by multiple classifiers. Green squares are AGN by MEx, brown squares are AGN by BPT, and yellow  triangles are AGN by WISE regardless of the classification by the other methods. The marginal plots on the X and Y axes, colour-coded using the same key, represent the density distribution of the points. AGN detected by \textsc{ProSpect} alone (blue) are distributed towards low host galaxy stellar mass. These AGN are also distributed towards low radio luminosity when compared to AGN detected by other diagnostics.}
    \label{fig:mass_lum}
\end{figure}

\noindent This could potentially explain why narrow-line diagnostics are relatively less efficient at detecting AGN in low mass host galaxies. Massive galaxies, where the black hole is a larger proportion by mass, can produce ionisation lines strong enough that then dominate and are easily detected by BPT and MEx. However, when the black holes are proportionally less massive, which is the case with low mass galaxies, the stellar contribution may outweigh the AGN contribution in the optical spectra, consequently minimising the AGN optical signature. This argument is reinforced by figure \ref{Dn4000} since the AGN identified by \textsc{ProSpect} alone lie in the LERG region, consistent with having weak optical emission features. A more detailed investigation of this relationship is explored by \cite{2018MNRAS.478.5074S}. That analysis links radio jets (and hence radio luminosity), accretion disks (optical emission line luminosity) and black hole mass. It is feasible for
lower-mass black holes to produce a radio jet that is more powerful (per unit black hole mass) than their higher-mass counterparts, especially if they are accreting close to the Eddington limit \citep{2015ApJ...806...59T}.

\noindent Figure \ref{fig:mass_lum} shows radio luminosity as a function of stellar mass colour coded by the AGN diagnostics. This is a summary of the key outcomes regarding the capabilities of different classifiers. The figure reiterates the capabilities of \textsc{ProSpect} in picking up potentially low radio luminosity AGN in low mass hosts over and above those picked up by other diagnostics.

\noindent Many AGN diagnostics exist, each with its own capabilities and parameter space in which they maximise their probability of identifying authentic AGN. The binary classifier scheme that we use makes it easy to see AGN uniquely identified by different diagnostic schemes. Not all diagnostics are perfectly coupled and the presence of AGN may not be simultaneously perceptible by different techniques. Since almost all galaxies are believed to host supermassive black holes, it seems likely that identifying their presence as an AGN becomes more probable as more AGN diagnostics are brought to bear. Thus we suggest that the better discussion to have is about the fraction of AGN and star formation contribution from a galaxy rather than as an artificial binary distinction into AGN or SFG. The probabilistic approach suggested by \cite{2011ApJ...736..104J} is an important step in this direction. The mixing between emission from star-formers and AGN in the BPT diagnostic have been studied as well \citep[e.g.,][]{2018ApJ...856...89T,2018ApJ...861L...2T} and now SED tools like \textsc{ProSpect} can add a new dimension to further extend this more detailed approach.
\section{Conclusion}\label{sec:conclusion}
\noindent We have presented a spectral energy distribution fitting technique as a tool to identify nuclear activity. We have used the GAMA DR4 G23 data (spectrophotometry and stellar masses), the EMU early science data, and the ALLWISE catalogue to perform a panchromatic SED fitting using \textsc{ProSpect}, a generative SED fitting package. We used four AGN diagnostic tools: the BPT diagram and the mass-excitation (MEx) diagnostic diagram (two narrow-line based methods), the WISE colour criterion, together with the SED-based \textsc{ProSpect} AGN identification tool. 

\noindent The combined sample of 2956 G23 field galaxies were classified by each of these four diagnostics. For ease of illustration, we introduce a binary-based identification scheme (table \ref{tab:bin}) where \verb|0| corresponds to SFGs and \verb|1| corresponds to AGN. The sample was divided into four mass bins and we found 81 low mass galaxies hosting AGN unique to \textsc{ProSpect} at masses below $10^{10}~M_\odot$. These low mass hosts are weaker in their radio emission as well compared to those identified by other methods. A cross-match with the low mass AGN hosts in GAMA G23 compiled by \cite{2022ApJ...937....7S} also confirms that the SED-based low mass AGN hosts are unique to \textsc{ProSpect}. Narrow-line diagnostics appear to struggle in identifying AGN in low mass galaxies. We present a possible cause for this selection difference in terms of the proportion of black hole mass in low and high mass galaxies. Low mass galaxies host proportionally lower mass black holes \citep{2023MNRAS.520.1975G} thereby enabling their star forming signature to dominate and resulting in these galaxies being identified as star formers by narrow-line diagnostics. This effect is likely to be more prominent for radio galaxies in their low-excitation mode.

\begin{acknowledgement}
\noindent We thank Philip Best for providing the necessary data points for the demarcation line in figure \ref{Dn4000}. JP is supported by the Australian Government International Research Training Program (iRTP) Scholarship. IP acknowledges support from CSIRO under its Distinguished Research Visitor Programme to work on the GAMA23 EMU Early Science project. JA acknowledges financial support from the Science and Technology Foundation (FCT, Portugal) through research grants PTDC/FISAST/29245/2017, UIDB/04434/2020 (DOI:10.5449\-9/UIDB/04434/2020) and UIDP/04434/2020 (DOI:10.54499/\-UIDP/04434/2020). MBi is supported by the Polish National Science Center through grants no. 2020/38/E/ST9/00395, 2018/30/E/ST9/00698, 2018\-/31/G/ST9/03388 and 2020/39/B/\-ST9/03494, and by the Polish Ministry of Science and Higher Education through grant DIR/WK/2018/12.

\noindent This scientific work uses data obtained from Inyarrimanha Ilgari Bundara / the Murchison Radio-astronomy Observatory. We acknowledge the Wajarri Yamaji People as the Traditional Owners and native title holders of the Observatory site. The Australian SKA Pathfinder is part of the Australia Telescope National Facility (\url{https://ror.org/05qajvd42}) which is managed by CSIRO. Operation of ASKAP is funded by the Australian Government with support from the National Collaborative Research Infrastructure Strategy. ASKAP uses the resources of the Pawsey Supercomputing Centre. Establishment of ASKAP, the Murchison Radio-astronomy Observatory and the Pawsey Supercomputing Centre are initiatives of the Australian Government, with support from the Government of Western Australia and the Science and Industry Endowment Fund. This paper includes archived data obtained through the CSIRO ASKAP Science Data Archive, CASDA (\url{http://data.csiro.au}).

\noindent GAMA is a joint European-Australasian project based around a spectroscopic campaign using the Anglo-Australian Telescope. The GAMA input catalogue is based on data taken from the Sloan Digital Sky Survey and the UKIRT Infrared Deep Sky Survey. Complementary imaging of the GAMA regions is being obtained by a number of independent survey programmes including {\em GALEX\/} MIS, VST KiDS, VISTA VIKING, {\em WISE\/}, {\em Herschel}-ATLAS, GMRT and ASKAP providing UV to radio coverage. GAMA is funded by the STFC (UK), the ARC (Australia), the AAO, and the participating institutions. The GAMA website is \url{http://www.gama-survey.org/}.
\end{acknowledgement}
\bibliography{References}
\end{document}